\documentclass{article}
\usepackage[utf8]{inputenc}
\usepackage{epsfig,latexsym,amsfonts,amsmath,amsthm,amssymb,amsbsy,multirow,slashed,wasysym,textcomp,comment,mathtools,bbold,cite,datetime}
\usepackage{geometry}
 \usepackage{pgfplots}
 \usepgfplotslibrary{patchplots}
 \pgfplotsset{compat=newest}
 
\usepackage{hyperref}
 \hypersetup{
      linktoc=all,    
  }

\usepackage{blindtext}

\usepackage{tikz}
\usetikzlibrary{arrows.meta}
\tikzstyle{bag} = [align=center]
\usetikzlibrary{decorations.pathmorphing}

\def\bz{{\bar z}}
\def\bw{{\bar w}}

 \newcommand{\badat}{\begin{alignedat}}
 \newcommand{\eadat}{\end{alignedat}}

\theoremstyle{definition}

\DeclareGraphicsExtensions{.pdf,.png}
\usetikzlibrary{patterns}
\usetikzlibrary{math}

\usetikzlibrary{arrows.meta}
\tikzstyle{bag} = [align=center]
\usetikzlibrary{decorations.pathmorphing}
\usepackage[normalem]{ulem}

\usetikzlibrary{calc}

\begin{document}
 \begin{titlepage}
 \begin{flushright}{${}$}
 \end{flushright}
	 \vspace{1.5cm}
\begin{center}
  \baselineskip=13pt 
  {\huge Chaos in Celestial CFT}
   \vskip1.5cm 
   {\large Sabrina Pasterski}${}^\diamondsuit$
   {\large and Herman Verlinde}${}^\blacklozenge$
  \\[8mm]
   {\em${}^\blacklozenge$ Physics Department, Joseph Henry Laboratories}\\[3mm]
\noindent{${}^\diamondsuit$ \it Princeton Center for Theoretical Science}\\[3mm]
{\em Princeton University, Princeton, NJ 08544, USA}

\end{center}

\vspace{2cm}

\begin{abstract}
\addtolength{\baselineskip}{.5mm}

Celestial holography proposes a duality between gravitational scattering in asymptotically flat space-time and a conformal field theory living on the celestial sphere. Its dictionary relates the infinite dimensional space-time symmetry group to Ward identities of the CFT. The spontaneous breaking of these asymptotic symmetries governs the dynamics of the soft sector in the CFT. Here we show that this sector encodes non-trivial backreaction effects that exhibit characteristics of maximal quantum chaos. A key element in the derivation is the identification of the Hilbert space of celestial CFT, defined through radial quantization, with that of a constantly accelerating Rindler observer.
From the point of view of the bulk, Rindler particles exhibit Lyapunov behavior due to shockwave interactions that shift the observer horizon. From the point of view of the boundary, the superrotation Goldstone modes affect the relevant representations of the celestial Virasoro symmetry in a manner that induces Lyapunov behavior of out-of-time-ordered celestial correlators.

\end{abstract}

\vfill

\end{titlepage}


\addtolength\baselineskip{1mm}
\tableofcontents

\setcounter{tocdepth}{3}

\addtolength\parskip{.5mm}
\addtolength\baselineskip{-.5mm}
\def\is{\! & \! = \! & \!}
\def\half{{1\over 2}}
\numberwithin{equation}{section}
\def\sub{\scriptscriptstyle}
\def\ip{${\mathcal I}^+$}
\def\calr{{\cal R}}
\def\e{{\epsilon}}
\def\g{{\gamma}}
\def\cs{{\cal S}}
\def\S{\Sigma }
\def\s{\sigma }
\def\sz{\sigma^{0} }
\def\Psz{\Psi^{0} }
 \def\p{\partial}
 \def\bz{{\bar z}}
 \def\cT{8\pi G_N {\mathcal T}}
\def\0{{(0)}}
\def\1{{(1)}}
\def\2{{(2)}}
 \def\cL{{\cal L}}
\def\co{{\cal O}}\def\cv{{\cal V}}
\def\n{\nabla}
\def\ci{{\mathcal I}}
\def\ipp{${\mathcal I}^+_+$}
\def\<{\langle }
\def\>{\rangle }
\def\[{\left[}
\def\]{\right]}
\def\bw{{\bar w}}
\def\h{{h^i_i}}
\def\t{{\rm{trace}}}
\def\o{\omega }
\def\ra{\bigr\rangle}
\newcommand{\non}{\nonumber}
\renewcommand{\O}{\Omega}
\renewcommand{\L}{\Lambda}
\newcommand{\bigO}{\mathcal{O}}
\newcommand{\sech}{\mbox{sech}}
\newcommand{\tn}{\tilde{n}}
\newcommand{\W}{\mathcal{W}}
\newcommand{\tr}{\mbox{tr}}
\newcommand{\Del}{\nabla}
\newcommand{\hs}[1]{\mbox{hs$[#1]$}}
\newcommand{\w}[1]{\mbox{$\W_\infty[#1]$}}
\newcommand{\bif}[2]{\small\left(\!\!\begin{array}{c}#1 \\#2\end{array}\!\!\right)}
\renewcommand{\u}{\mathfrak{u}}
\newcommand{\scriplus}{\mathcal{I}^+}
\renewcommand{\epsilon}{\varepsilon}
\def\dim#1{\lbrack\!\lbrack #1 \rbrack\!\rbrack }
\newcommand{\chichi}{\chi\!\cdot\!\chi}
\newcommand\snote[1]{\textcolor{magenta}{[S:\,#1]}}

\renewcommand{\theequation}{\thesection.\arabic{equation}}
   \makeatletter
  \let\over=\@@over \let\overwithdelims=\@@overwithdelims
  \let\atop=\@@atop \let\atopwithdelims=\@@atopwithdelims
  \let\above=\@@above \let\abovewithdelims=\@@abovewithdelims
\renewcommand\section{\@startsection {section}{1}{\z@}%
                                   {-3.5ex \@plus -1ex \@minus -.2ex}
                                   {2.3ex \@plus.2ex}%
                                   {\normalfont\large\bfseries}}

\renewcommand\subsection{\@startsection{subsection}{2}{\z@}%
                                     {-3.25ex\@plus -1ex \@minus -.2ex}%
                                     {1.5ex \@plus .2ex}%
                                     {\normalfont\bfseries}}

\newcommand{\Tr}{\mbox{Tr}}
\renewcommand{\H}{\mathcal{H}}
\newcommand{\SU}{\mbox{SU}}
\newcommand{\chiu}{\chi^{{\rm U}(\infty)}}
\newcommand{\ff}{\rm f}
\linespread{1.3}

\def\bfR{{\mbox{\textbf R}}}
\def\gzz{\gamma_{z\bz}}
\def\vx{{\vec x}}
\def\p{\partial}
\def\po{$\cal P_O$}
\def\cN{{\cal H}^+ }
\def\N{${\cal H}^+  ~~$}
\def\G{\Gamma}
\def\l{{\ell}}
\def\ch{{\cal H}^+}
\def\Q{{\hat Q}}
\def\T{\hat T}
\def\C{\hat C}
\def\zet{z}
\def\scc{\mbox{\small $\hat{C}$}}
\def\Aa{{\mbox{\scriptsize \smpc \sc a}}}
\def\aalpha{{\mbox{\scriptsize {\smpc  $\alpha$}}}}
\def\cC{{\mbox{\scriptsize {\smpc \sc c}}}}
\def\cCb{{\mbox{\scriptsize \smpc \sc c'}}}
\def\sS{{\mbox{\scriptsize {\smpc \sc s}}}}
\def\Bb{{\mbox{\scriptsize \smpc \sc b}}}
\def\Hh{{\mbox{\scriptsize \smpc \sc h}}}
\def\oO{{}} 
\def\bfC{\mbox{{\textbf C}}}
\def\nonu{\nonumber}
\def\im{{\rm i}}
\def\tr{{\rm tr}}
\def\be{\bea}
\def\ee{\eea}

\def\spc{\hspace{.5pt}}

\def\bea{\begin{eqnarray}}
\def\eea{\end{eqnarray}}
\def\half{{\textstyle{\frac 12}}}
\def\cL{{\cal L}}
\def\halfi{{\textstyle{\frac i 2}}}

\def\delbar{\overline{\partial}}
\newcommand{\smpc}{\hspace{.5pt}}
\def\nspc{\!\spc\smpc}
\def\uU{\mbox{\textit{\textbf{U}}\spc}}
\def\uV{\mbox{\textit{\textbf{U}\spc}}}
\def\tT{\mbox{\textit{\textbf{T}}\spc}}
\def\bfC{\mbox{\textit{\textbf{C}}}}
\def\bn{\mbox{\textit{\textbf{n}\!\,}}}
\def\pP{\mbox{\textit{\textbf{P}\!\,}}}
\def\rR{{\textit{\textbf{R}\!\,}}}

\def\im{{\rm i}}
\def\tr{{\rm tr}}

\def\ra{\bigr\rangle}
\def\la{\bigl\langle}
\def\li{\bigl |\spc}
\def\ri{\bigr |\spc}

\def\nonu{\nonumber}
\def\SL2{SL(2,\mathbb{R})}
\def\mR{\mathbb{R}}
\def\mZ{\mathbb{Z}}
\def\nn{\nonumber}

\def\centerarc[#1](#2)(#3:#4:#5)
    { \draw[#1] ($(#2)+({#5*cos(#3)},{#5*sin(#3)})$) arc (#3:#4:#5); }

\enlargethispage{\baselineskip}

\setcounter{tocdepth}{2}
\newpage 
\addtolength{\baselineskip}{.3mm}
\addtolength{\parskip}{.3mm}
\addtolength{\abovedisplayskip}{.9mm}
\addtolength{\belowdisplayskip}{.9mm}
\renewcommand\Large{\fontsize{15.5}{16}\selectfont}

\newcommand{\newsubsection}[1]{
\vspace{.6cm}
\pagebreak[3]
\addtocounter{subsubsection}{1}
 \addcontentsline{toc}{subsection}{\protect
 \numberline{\arabic{section}.\arabic{subsection}.\arabic{subsubsection}}{#1}}
\noindent{\arabic{subsubsection}. \bf #1}
\nopagebreak
\vspace{1mm}
\nopagebreak}
\renewcommand{\footnotesize}{\small}

\section{Introduction}
\vspace{-2mm}

Celestial conformal field theory (CCFT) aims to provide a holographic dual description of four-dimensional quantum gravity in asymptotically flat space-time~\cite{Strominger:2017zoo,Raclariu:2021zjz,Pasterski:2021rjz}.   Its dictionary exploits the fact that the 4D Lorentz group SL$(2,\mathbb{C})$ acts via two-dimensional global conformal transformations on the celestial sphere~$\mathbb{S}^2$ and postulates an identification between 4D scattering amplitudes and correlation functions of local operators in a putative 2D CFT defined on~$\mathbb{S}^2$~\cite{deBoer:2003vf,Strominger:2013jfa,Strominger:2013lka,Cheung:2016iub,Pasterski:2016qvg,Pasterski:2017kqt,Pasterski:2017ylz}.  A key feature of this framework is that it prioritizes infinite dimensional symmetry enhancements associated to the asymptotic symmetry group in the bulk.  Via the identification between amplitudes and correlation functions, soft theorems of 4D quantum gravity translate into an infinite set of conformal Ward identities in the 2D dual.  

To set notation, let us briefly summarize the celestial holographic mapping. Consider a massless scattering amplitude $A(p_i)$ in four-dimensional asymptotically flat space-time as a function of the on-shell momenta $p_i$ for the external scattering states. A lightlike momentum vector $p^\mu$ is parametrized by a direction $(z,\bz)$ on the celestial sphere $ \mathbb{S}^2$ and a light-cone momentum $\omega$ via 
\bea
\label{ptosphere}
p^\mu = \pm\omega q^\mu , \qquad q^\mu \! \is \! \frac1 2  \bigl( 1\nspc +\nspc  \bz z, z \nspc +\nspc  \bz, i (\bz\nspc - \nspc z), 1\nspc -\nspc z\bz \bigr).
\eea
The basis change between the amplitude  $A(p_i)\! = \! A(\omega_i; z_i, \bz_i)$ in the momentum eigenbasis and the amplitude in the boost eigenbasis proceeds via a Mellin transform 
\bea\label{Mellin}
A(\Delta_i, z_i,\bz_i) \is \Bigl[ \, \prod_i \int_0^\infty\!\! d\omega_i\, \omega_i^{\Delta_i-1} \Bigr] 
A(\omega_i;z_i,\bz_i)\,.
\eea
Because the corresponding external wavefunctions transform covariantly under SL$(2,\mathbb{C}$), the Mellin amplitudes behave like conformal correlation functions of a local 2D CFT. This motivates the identification 
\bea\label{Arescaled}
A(\Delta_i,z_i,\bz_i) \! \is \! {\cal N} \, \Bigl\langle \mathcal{O}^{\pm}_{\Delta_1}(z_1,\bz_1)\ldots \mathcal{O}^{\pm}_{\Delta_n}(z_n,\bz_n) \Bigr\rangle
\eea
where ${\cal N} = {\prod_k \; i^{\mp \Delta_k } \Gamma(\Delta_k) }$ and $\mathcal{O}^{\pm}_{\Delta}(z,\bz)$ denote local primary operators of a putative celestial CFT.   The $\pm$ phase depends on whether the particle is incoming or outgoing.

While CCFT has proven to be an effective framework for codifying the infrared symmetry properties of 4D scattering amplitudes~\cite{Strominger:2017zoo}, little is known about its dynamics, nor has it yet been employed to make dynamical predictions about 4D quantum gravity other than those that follow from symmetries. The main obstacle towards extracting such dynamical predictions is that an intrinsic construction of celestial CFT starting from a microscopic theory of quantum gravity is still lacking. It may thus seem premature to conclude that CCFT represents a conventional local QFT. Indeed, celestial correlation functions share some but not all properties of standard 2D CFT correlation functions. However, a first indication that celestial CCFT has local dynamics is that it has a candidate local stress energy tensor in the form of the subleading soft graviton mode~\cite{Kapec:2016jld}.  We will find that it is still fruitful to adopt the viewpoint that CCFT exists as true physical quantum system and then use the holographic dictionary to deduce its dynamical properties. 

In AdS holography, gravitational shockwave dynamics in the vicinity of a black hole horizon is now understood to be a manifestation of chaotic quantum dynamics of the dual CFT, and vice versa~\cite{Shenker:2013pqa,Shenker:2013yza,Susskind:2013aaa,Jackson:2014nla,Roberts:2014ifa}. The Lyapunov behavior of the CFT is governed by an emergent Goldstone mode associated with the breaking of translation invariance due to presence of the horizon, or on the CFT side, due to the presence of the thermal CFT plasma. 
At first sight, one would not expect celestial CFT to exhibit the same type of chaotic behavior: flat space-time does not possess intrinsic event horizons that lead to Lyapunov growth. However, since celestial holography trades manifest space-time translation symmetry for boost invariance, and boosts are defined relative to a choice of origin in 4D space-time, the celestial sphere should be thought of as being located at some specified lightcone time $u$ along null infinity. While boosts act linearly in CCFT, space-time translations are non-linearly realized: the translation generator along the $u$ direction acts on conformal primary fields by shifting the conformal dimension from $\Delta$ to $\Delta + 1$~\cite{Donnay:2018neh,Stieberger:2018onx}. 
This shift in conformal dimension is a first hint of exponential Lyapunov behavior, and indeed a first hint that from the CCFT point of view, 4D translation symmetry, rather than being an exact symmetry with associated conserved currents, should perhaps be thought as an emergent symmetry arising from underlying strongly coupled quantum dynamics.

Here we set out to study CCFT following the same logic used to exhibit chaotic dynamics in AdS.  We will proceed via a combination of three methods. First, we will identify the relevant dynamics of the conformally soft sector of the 4D gravity theory, as determined by the spontaneous symmetry breaking of the asymptotic symmetry group. Specifically, we will focus on the backreaction associated to the superrotation Goldstone modes.  We will see that, even without a full understanding of the holographic dictionary, we can make concrete statements about the holographic dictionary associated to the soft dynamics, and that this scope is naturally adapted to detect signals of chaotic phenomena.\footnote{As explored in~\cite{Pasterski:2020xvn}, the phase space of Goldstone modes for asymptotic symmetries captures backreaction effects due to matter in the vicinity of the event horizon.
While supertranslations correspond to the leading soft graviton theorem~\cite{He:2014laa}, the subleading soft graviton theorem plays a more intrinsic role in CCFT~\cite{Cachazo:2014fwa,Kapec:2014opa,Pasterski:2019ceq}.} 

\begin{figure}[t]
\begin{center}
\begin{tikzpicture}[scale=.95]
\draw[gray]    (-3.3,-1.6) to[out=90,in=-155] (-2,0.275);
\draw[gray]    (-3.3,-1.6) to[out=-90,in=155] (-2,-3.275);
\draw[gray]    (-.7,-1.6) to[out=90,in=-25] (-2,0.275);
\draw[gray]    (-.7,-1.6) to[out=-90,in=25] (-2,-3.275);
\draw[gray]    (-2.7,-1.7) to[out=90,in=-130] (-2,0.25);
\draw[gray]    (-2.7,-1.7) to[out=-90,in=130] (-2,-3.25);
\draw[gray]    (-1.3,-1.7) to[out=90,in=-50] (-2,0.25);
\draw[gray]    (-1.3,-1.7) to[out=-90,in=50] (-2,-3.25);
\draw[gray]    (-2,-1.7) to[out=90,in=-90] (-2,0.25);
\draw[gray]    (-2,-1.7) to[out=-90,in=90] (-2,-3.25);
\draw[yscale=.4,->] (-2.5,-6.2) arc (-135:-45:.75) node[above left]{\raisebox{-4mm}{$\phi$}};
\draw[xscale=.4,->] (0,-2) arc (-40:40:.7) node[below right]{\raisebox{-4mm}{\ $\tau$}};
\filldraw[black, thick] (-2,-3.25) circle (.2em);
\filldraw[black, thick] (-2,0.25) circle (.2em);
\draw[black, thick] (-2,-1.5) circle (5em);
\draw[dashed,yscale=.25] (-2,-6) circle (5em);
\draw[thick,yscale=.4] (3.666,0) circle (1.9em);
\draw[thick] (3,0) -- (3,-3);
\draw[gray] (3.2,-0.18) -- (3.2,-3.18);
\draw[gray] (3.433,-0.25) -- (3.433,-3.25);
\draw[gray] (3.666,-0.28) -- (3.666,-3.28);
\draw[gray] (3.9,-0.25) -- (3.9,-3.25);
\draw[gray] (4.133,-0.18) -- (4.133,-3.18);
\draw[thick] (4.33,0) -- (4.33,-3);
\draw[thick,yscale=.4] (3.666,-7.5) circle (1.9em);
\draw[dashed,yscale=.4] (3.666,-3.75) circle (1.9em);
\draw[yscale=.5,->] (3.3,-5) arc (-135:-45:.5) node[above left]{\raisebox{-4mm}{$\phi$}};
\draw[->] (4.75,-2) -- (4.75,-1)  node[below right]{\raisebox{-4mm}{$\tau$}};
\end{tikzpicture}
\end{center}
\caption{The  cylinder coordinates $\tau$ and $\phi$ defined through \eqref{expmap} map the celestial sphere onto the celestial cylinder. We will argue that the CCFT dynamics in the $\tau$ direction exhibits maximal quantum chaos.}
\label{fig:tauphi}
\end{figure}
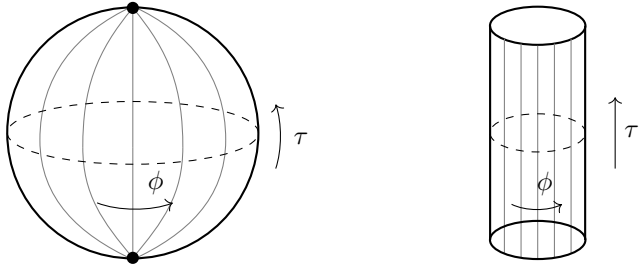

Second, taking a more geometrical perspective, we will argue that the CCFT Hilbert space defined through radial quantization should be identified with the Hilbert space of a Rindler observer following a constantly accelerating trajectory that reaches asymptotic infinity at the pole of the celestial sphere. Standard radial quantization involves a mapping to the celestial cylinder via the exponential coordinate transformation
\bea\label{expmap}
z= e^{-\tau + i\phi}, \quad & &\quad \bz = e^{-\tau-i\phi}.
\eea
The $\phi$ and $\tau$ evolution are indicated in figure~\ref{fig:tauphi}. Our proposal is that the CCFT dynamics in the $\tau$ direction in fact takes place at finite temperature $T$ and, moreover, displays quantum chaos with a Lyapunov exponent that saturates the chaos bound.

The conclusion that CCFT has finite temperature naturally follows by requiring that celestial amplitudes arise from analytic continuation to (2,2) signature. This motivates the interpretation of the CCFT Hilbert states as describing the quantum state of the Rindler horizon. The observations of a Rindler observer are restricted to a wedge of space-time outside event horizon and as a consequence, she experiences her environment as in a mixed quantum state with finite temperature. Moreover, to this observer, the shockwave interaction due to an incoming particle crossing the horizon will induce Lyapunov type behavior, similar to what happens for black hole horizons. Extrapolating lessons learned from AdS/CFT, one is led to conclude that the Rindler horizon represents a strongly coupled physical quantum system with maximal chaos. Via our working assumption that celestial holography represents a true duality between two physical systems, this implies that the CCFT should also exhibit maximal chaos.  Finally, combining these insights with well-established general properties of correlation functions and conformal blocks in 2D CFT, we will study the out-of-time-ordered correlation functions (OTOCs) in CCFT and argue that they indeed display signatures of maximally chaotic dynamics.

\begin{figure}[t]
\centering
\vspace{-0.5em}
\begin{tikzpicture}[scale=2.7]
\definecolor{darkgreen}{rgb}{.0, 0.5, .1};
\draw[thick](0,0) --(1,1) node[right] {$i^0$} --(0,2)node[above] {$i^+$} --(-1,1) --(0,0)  node[below] {$i^-$} ;
\draw[dashed] (-1/2,1/2) --(1/2,3/2) node[above right] {$\cal{I}^+$};
\draw[dashed] (1/2,1/2) node[below right] {$\cal{I}^-$} --(-1/2,3/2);
\draw[very thick,darkgray,dashed,-latex] (1/2+.01,1/2+.01) to [bend left=45] (1/2+.015,3/2-.015);
\draw[darkgray] (0.14,1) node[right] {$R$};
\draw[blue] (-.3,.65) node[right] {$A$};
\draw[red] (.65,.9) node[above] {$B$};
\draw[thick,red,-latex] (.75,.75) -- (-.25,1.75);
\draw[thick,blue]  (-.475,.475) -- (.275,1.225);
\draw[thick,blue,-latex] (.225,1.275) -- (.475,1.525);
\end{tikzpicture}
\vspace{-3mm}
\caption{Radial time evolution on the celestial sphere maps to time evolution of Rindler observer in space-time. A particle experiences a Shapiro time delay when crossing a shockwave. From the perspective of a fiducial Rindler observer, this time delay may have drastic effect that grows exponentially in time.}
\label{shockshift}
\end{figure}
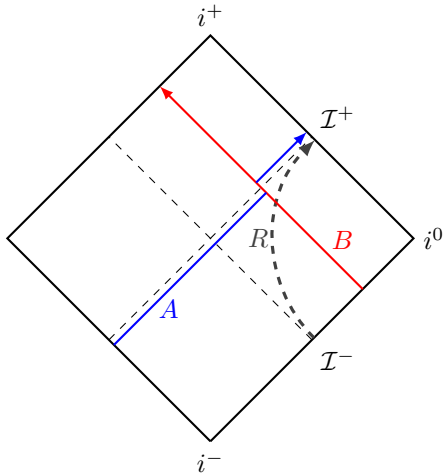

This paper is organized as follows. We review the celestial dictionary and the emergence of celestial Virasoro symmetry in section~\ref{sec:dictionary}.  We then study the celestial backreaction in section~\ref{sec:vir} and relate the associated Goldstone mode dynamics with properties of the CCFT. In particular, we will make a concrete proposal for the central charge of the Virasoro symmetry of CCFT. We then proceed to study the bulk interpretation of radial quantization in section~\ref{sec:ccftrindler} and show how to change between signatures in both the bulk and boundary. Finally, with all these ingredients in place, we then study CCFT on the celestial torus in section~\ref{CCFTorus} and demonstrate the chaotic behavior of OTOCs in section~\ref{OTOCinCCFT}.

\section{Celestial dictionary and symmetries}\label{sec:dictionary}

\vspace{-1mm}

To set the stage, we start with a brief review of some basic elements of celestial CFT including the bulk-to-boundary mapping from local operators in 4D space-time to local CFT operators on the 2D celestial sphere and the identification of an infinite asymptotic symmetry algebra isomorphic to the Virasoro algebra. 

\subsection{Bulk to boundary mapping} 

\vspace{-1mm}

A {\it conformal primary wavefunction} is a wavefunction $\Phi_{\Delta,J}(X^\mu;z,\bz)$ on $\mathbb{R}^{1,3}$ which transforms under SL$(2,\mathbb{C})$ as a 4D tensor field of spin-$s$ and as a 2D conformal primary of conformal dimension $\Delta$ and spin $J$
\be\label{deltaj2}
\Phi_{\Delta,J}\Big(\Lambda^{\mu}_{~\nu} X^\nu;\frac{a z+b}{cz+d},\frac{{\bar a} \bz+{\bar b}}{{\bar c}\bz+{\bar d}}\Big)=(cz+d)^{\Delta+J}({\bar c}\bz+{\bar d})^{\Delta-J}D(\Lambda)\Phi_{\Delta,J}(X^\mu;z,\bz)\,,\\[-2mm] \nonumber
\ee
where $D(\Lambda)$ is the 3+1D spin-$s$ representation of the Lorentz algebra~\cite{Pasterski:2016qvg,Pasterski:2017kqt,Pasterski:2020pdk}. A special subclass of conformal primary wave functions are the perturbative  wave functions with that solve the linearized equations of motion for a massless (in which case $J=\pm s$) or massive spin-$s$ particle in vacuum. 
Given a local 4D operator $\hat{O}(X)$ and a linearized conformal primary wavefunction $\Phi^\pm_{\Delta,J}(X;z,\bz)$, we can define the associated local CCFT operator $\mathcal{O}^\pm_{\Delta,J}(z,\bz)$ by taking the overlap computed via the appropriate Klein-Gordon inner product $\bigl(\; ,\; \bigr)_{\Sigma}$ on a Cauchy slice $\Sigma$ 
\bea
\label{kgin}
\mathcal{O}^\pm_{\Delta,J}(z,\bz)\! \is \! i\bigl(\, \hat{O}\spc ,\spc \Phi^\pm_{\Delta,J}(z,\bz)^*\bigr)_{\Sigma}
\eea
with $\Phi^\pm_{\Delta,J}(X;z,\bz)^* = \Phi^\mp_{\Delta^*,-J}(X;z,\bz)$.
Here the $\pm$ superscript indicates whether the operator prepares an outgoing or incoming state, respectively.  In the interacting theory we want to push the Cauchy slice to future or past null infinity to prepare the out or in states from the corresponding 4D vacuum state.  

In this paper we will aim to go beyond perturbation theory and determine the properties of celestial correlation functions of general local operators that follow from universal gravitational dynamics in the bulk. The above perturbative dictionary, however, still gives useful guidance.  A first lesson is that, as  shown in \cite{Pasterski:2017kqt}, finite energy perturbative modes are captured by conformal dimensions on the principal series
\bea
\Delta\!\is \! 1+i\lambda,~~~\lambda\in \mathbb{R}.
\eea
Hence it is reasonable to conclude that the spectrum of primary operators in CCFT will in general include operators with complex conformal dimensions. This in particular has consequences for the Hermiticity properties of the CCFT Hamiltonian and for the relation between celestial and bulk time-evolution. Secondly, the fact that there exists a map from bulk operators to boundary operators gives a direct motivation for our working hypothesis that the CCFT Hilbert space and bulk Hilbert space should be identified. This also means that the CCFT and bulk dynamics should be related. We will present a proposal for including dynamical time evolution in the bulk-to-boundary dictionary in section~\ref{sec:rad}.

In conventional treatments of celestial conformal field theory 4D translations are realized as an exact symmetry generated by conserved translation charges $P_\mu$ that act on the conformal primary operators via~\cite{Stieberger:2018onx}
\bea
P_\mu \mathcal{O}^\pm_{\Delta,J}(z,\bz)\is \pm q_\mu \mathcal{O}^\pm_{\Delta+1,J}(z,\bz).
\eea
The associated Ward identities imply the existence of infinite towers of primary operators with precise relationships between their OPE coefficients.
When applying standard celestial CFT technology we need to keep in mind these larger multiplet structures~\cite{Stieberger:2018onx,Law:2019glh,Banerjee:2020kaa} and consider the behavior of correlators in the complex $\Delta$ plane~\cite{Donnay:2020guq,Arkani-Hamed:2020gyp}.  Following this symmetry guided approach to its logical conclusion reveals that CCFT exhibits a large extended symmetry group in the form of a $w_{1+\infty}$-algebra \cite{Strominger:2021lvk}.  

In this paper we will largely ignore the enhanced symmetry structure of CCFT, including those that follow from 4D translation symmetry. Instead we will mostly concentrate on properties of that follow from 4D Lorentz symmetry and its infinite dimensional extension, superrotation symmetry.

\subsection{Asymptotic symmetry algebra} \label{sec:asg}

\vspace{-1.5mm} 

The key evidence that underlies celestial conformal field theory is that the asymptotic superrotation symmetries of gravity  act on the celestial sphere via local 2D conformal transformations.  This implies the existence of an infinite set of superrotation generators, which can be combined to constitute the local stress energy tensor of the CCFT. We start with a brief review of this result, while highlighting the role of the conformally soft modes.

In Bondi gauge, the metric near future null infinity takes the form~\cite{Bondi:1962px,Sachs:1962wk} 
\be\badat{3}\label{bondi}
ds^2=&-du^2-2dudr+2r^2\gamma_{z\bz}dzd\bz+\frac{2m_B}{r}du^2+rC_{zz}dz^2+rC_{\bz\bz}d\bz^2\\
&+\bigl[(D^zC_{zz}-\frac{1}{4r}D_z(C_{zz}C^{zz})+\frac{4}{3r}(N_z+u\p_z m_B))dudz+c.c.\bigr]+...
\eadat\ee
using the conventions of~\cite{Strominger:2017zoo}. The radiative data are captured by the news tensor $N_{zz}=\p_u C_{zz}$, which can be specified as a free function of $(u,z,\bz)$ while the $u$ evolution of the Bondi mass $m_B$ and angular momentum aspect $N_z$ are determined by the constraint equations $G_{u\mu}=8\pi G T^M_{u\mu}$ at large-$r$.

The asymptotic symmetry group is determined by identifying residual diffeomorphisms which preserve these falloffs but act non-trivially near $\mathcal{I}^+$.  These include supertranslations which shift the generators of null infinity by a free function of $(z,\bz)$, as well as superrotations which enhance the Lorentz subgroup and will be our focus here.  In Bondi gauge the superrotation vector fields take the form
\bea\label{xiy}
\xi_Y\!\!\is\!\! (1+\frac{u}{2r})Y\p_z-\frac{u}{2r}D^\bz D_z Y\p_\bz-\frac{1}{2}(u+r)D_zY\p_r +\frac{u}{2}D_zY\p_u+c.c.
\eea
Under these diffeomorphisms, the news transforms as 
\bea
\delta_Y N_{zz}\!\!\is\!\!\frac{u}{2}D_A Y^A\p_u N_{zz}+\mathcal{L}_Y N_{zz}-D_z^3 Y^z.
\eea
The last term is an inhomogeneous shift of the superrotation Goldstone mode~\cite{Himwich:2019qmj,Ball:2019atb}. 

A key observation by Strominger~\cite{Strominger:2017zoo} is that Ward identities for asymptotic symmetries are manifested as soft theorems of the $\mathcal{S}$-matrix. The canonical charge generating the superrotation symmetry can be split into a soft and a hard part~\cite{Kapec:2014opa}. Combining the contributions from future $(+)$ and past $(-)$ null infinity (or more generally a signed sum over boundary components) we get a Ward identity of the form
\be\label{ward}
Q^\pm=Q^\pm_S+Q^\pm_H,~~~Q=Q^+-Q^-=0.
\ee
  Acting on the vacuum, the operator $Q_S$ adds a soft graviton mode, while $Q_H$ induces the corresponding symmetry transformation of the matter fields and finite energy gravitational fluctuations. 
  
The real power of the celestial CFT framework derives from the identification of the subleading soft graviton with the local 2D stress energy tensor~\cite{Kapec:2016jld} 
\bea\label{qs}
T_{zz}^{\rm\spc CFT}\!\!\nspc \is\nspc \! \frac{1}{8\pi G}\int\!\nspc du \spc u \widetilde{N}_{zz}. \quad 
\eea
Here $ \widetilde{N}_{zz} \equiv -6 i  \int\! \nspc d^2 w \sqrt{\gamma} \frac 1 {(z-w)^4} N^{ww}$ amounts to
a shadow transform of the news tensor.\footnote{This is a proper 2D conformal shadow transform if the $u$-integral is performed first. Namely, the $duu$-integral of the news gives a weight $\Delta=0$, $J=-2$ operator whose shadow has weight $\Delta=2$, $J=2$.} This operator
inserts a zero energy graviton that couples to angular momentum and boost energy. Its charges 
\bea
Q_{S}(Y) \! \is \! 
\oint \frac{dz}{2\pi i}Y^zT^{\rm CFT}_{zz}
\eea
generate the superrotation transformations and
measure the spin memory effect~\cite{Pasterski:2015tva}.  

Superrotations act on the celestial sphere as  local conformal transformations $z \to z + Y(z)$. The Laurent modes of the stress tensor thus generate a Virasoro algebra
\bea
\label{virone}
T_n=\oint\! \frac{dz}{2\pi i}\spc z^{n+1} T_{zz}^{\rm CFT}, \quad \qquad [T_{n},T_m]\!\!\is\!\! (n\nspc -\nspc m)T_{n+m}
\eea
with vanishing central charge.
The insertion of a stress tensor in a CCFT correlation function of primary local operators gives rise to the familiar conformal Ward identity of 2D CFT
\bea\label{Tward}
& &\bigl\langle\, T_{zz}^{\rm CFT}(z) \, \mathcal{O}_1(z_1)\ldots \mathcal{O}_n(z_n)\bigr\rangle=\sum_k\left[\frac{h_{k}}{(z-z_k)^2}+\frac{\p_{z_k}}{z-z_k}\right]\bigl\langle\mathcal{O}_1(z_1) \ldots \mathcal{O}_n(z_n)\bigr\rangle
\eea
and similar for $\bar{T}_{\bz\bz}$.  In terms of our holographic dictionary, this 2D Ward identity follows from the subleading soft graviton theorem in 4D~\cite{Cachazo:2014fwa,Kapec:2014opa}.

\subsection{Superrotation Goldstone modes}

\vspace{-1.5mm}

Besides the CCFT stress tensor, the gravity theory contains geometric soft modes associated with supertranslations and  superrotations. In the absence of matter stress energy, the off-diagonal asymptotic metric component $C_{zz}$ can be decomposed as~\cite{Compere:2016jwb,Compere:2018ylh}
\bea
\label{cdeco}
C_{zz} \! \is \! (u+\mathcal{C})\Theta_{zz}-2D_z^2\mathcal{C}.
\eea
Here $\mathcal{C}$ is the supertranslation Goldstone mode and $\Theta_{zz}$ is the superrotation Goldstone mode. 

The $\Theta_{zz}$ mode encodes the celestial backreaction due to insertion of a classical source that couples to CCFT stress tensor $T_{zz}^{\it CFT}$. It transforms under superrotations~as (here $Y=Y^z$)\footnote{As compared to ex.~\cite{Kapec:2014opa} where  the canonical commutation relations in 4D imply $\delta_Y\Theta_{zz}= i[Q_S(Y),\Theta_{zz}]$ , our use of $\[~,~\]$ will denote the 4D Poisson bracket. This is intended to be suggestive for what follows since the algebra matches what is expected for the radial quantization commutator.
}
\bea
\label{thetatrafo}
\delta_Y\Theta_{zz}\!\! \is [Q_S(Y),\Theta_{zz}]\, =\, Y \p_z \Theta_{zz}+ 2 \p_z Y \Theta_{zz} - \p^3_zY.
\eea
We see that superrotation transformation rule of $\Theta_{zz}$ is identical to the conformal transformation rule of a 2D CFT stress tensor with unit central charge. Note, however, that $\Theta_{zz}$ itself is not a superrotation generator: its Laurent coefficients do not generate a Virasoro algebra but commute among themselves
\bea
\label{virzero}
\Theta_n\!\! \is \!\! \oint\! \frac{dz}{2\pi i} \spc z^{n+1} \Theta_{zz}
 \qquad \qquad [\Theta_{n},\Theta_m] \, =\, 0.
\eea
Meanwhile, the $T_n$ generators and $\Theta_m$ modes do satisfy a non-trivial commutator algebra. 
Equation \eqref{thetatrafo} implies that
\bea
\label{virtwo}
[T_n,\Theta_m] \! \is \! (n\nspc -\nspc m)\Theta_{n+m}\nspc -\nspc 
(m^3\!-m)\delta_{n+m}.
\eea
As we will show in the next section, the commutation relations \eqref{virone}, \eqref{virzero} and \eqref{virtwo} can be repackaged in the form of (an In\"on\"u-Wigner contraction of) a pair of Virasoro algebras with divergent central charge. 

The interpretation of $\Theta_{zz}$ as the superrotation Goldstone mode is made more manifest by writing it as
\bea
\label{thetaf}
\Theta_{zz}\! \is \! -\{Z(z),z\} 
\eea 
where $\{f, z\} = \frac{f'''}{f'}\nspc - \nspc \frac 3 2 \bigl(\frac{f''}{f'}\bigr)^2$ denotes the Schwarzian derivative. From equation \eqref{thetatrafo}, we then see that infinitesimal superrotations generate a linear shift in $Z(z)$ via
\bea
\delta_Y Z(z) \!\! \is [Q_S(Y),Z(z)] \, = \, Y(z) .
\eea
Hence, by exponentiation, $Z(z)$ represents the full non-linear superrotation Goldstone mode. The soft superrotation dynamics and the associated appearance of a Virasoro algebra with non-zero effective central charge will play a key role in the derivation of the Lyapunov behavior of CCFT.

\section{Celestial backreaction}\label{sec:vir}
\vspace{-1mm}

\begin{figure}[t]
\centering
\raisebox{-0cm}{
\begin{tikzpicture}[xscale=-.9,yscale=.9]
\draw[thick] (3,0) node[left]{\small $i^0$}-- (0,3); 
\draw[black!50!white] (0,0) --(1.5,1.5);
\draw[blue!50!white]    (0,2) to[out=0,in=130] (1.5,1.5);
\draw[red!50!white]    (2,0) to[out=90,in=-50] (1.5,1.5);
\draw[blue!50!white]    (0,1.5) -- (1.5,1.5);
\draw[red!50!white]    (1.5,0) -- (1.5,1.5);
\draw[blue!50!white]    (0,1) to[out=0,in=-140] (1.5,1.5);
\draw[red!50!white]    (1,0) to[out=90,in=-130] (1.5,1.5);
\draw[blue!50!white]    (0,.5) to[out=5,in=-140] (1.5,1.5);
\draw[red!50!white]    (.5,0) to[out=85,in=230] (1.5,1.5);
\draw[red!50!white]    (2.5,0)  to[out=95,in=-50] (1.5,1.5);
\draw[blue!50!white]    (0,2.5) to[out=-5,in=140] (1.5,1.5);
\end{tikzpicture}}
\hspace{-5.5mm}
\begin{tikzpicture}[scale=.9]
\draw[thick] (3,0) node[right]{\small $i^0$}-- (0,3) node[above]{\small $i^+$} ;
\draw[black!50!white] (0,0) --(1.5,1.5);
\draw[blue!50!white]    (0,2) to[out=0,in=130] (1.5,1.5);
\draw[red!50!white]    (2,0) to[out=90,in=-50] (1.5,1.5);
\draw[blue!50!white]    (0,1.5) -- (1.5,1.5);
\draw[red!50!white]    (1.5,0) -- (1.5,1.5);
\draw[blue!50!white]    (0,1) to[out=0,in=-140] (1.5,1.5);
\draw[red!50!white]    (1,0) to[out=90,in=-130] (1.5,1.5);
\draw[blue!50!white]    (0,.5) to[out=5,in=-140] (1.5,1.5);
\draw[red!50!white]    (.5,0) to[out=85,in=230] (1.5,1.5);
\draw[red!50!white]    (2.5,0)  to[out=95,in=-50] (1.5,1.5);
\draw[blue!50!white]    (0,2.5) to[out=-5,in=140] (1.5,1.5);
\end{tikzpicture}
\hspace{-6.25cm}
\raisebox{-2.445cm}{
\begin{tikzpicture}[xscale=-.9,yscale=-.9]
\draw[thick] (3,0) -- (0,3); 
\draw[black!50!white] (0,0) --(1.5,1.5);
\draw[blue!50!white]    (0,2) to[out=0,in=130] (1.5,1.5);
\draw[red!50!white]    (2,0) to[out=90,in=-50] (1.5,1.5);
\draw[blue!50!white]    (0,1.5) -- (1.5,1.5);
\draw[red!50!white]    (1.5,0) -- (1.5,1.5);
\draw[blue!50!white]    (0,1) to[out=0,in=-140] (1.5,1.5);
\draw[red!50!white]    (1,0) to[out=90,in=-130] (1.5,1.5);
\draw[blue!50!white]    (0,.5) to[out=5,in=-140] (1.5,1.5);
\draw[red!50!white]    (.5,0) to[out=85,in=230] (1.5,1.5);
\draw[red!50!white]    (2.5,0)  to[out=95,in=-50] (1.5,1.5);
\draw[blue!50!white]    (0,2.5) to[out=-5,in=140] (1.5,1.5);
\end{tikzpicture}}
\hspace{-6.675mm}
\raisebox{-2.95cm}{
\begin{tikzpicture}[xscale=.9,yscale=-.9]
\draw[thick] (3,0) -- (0,3) node[below]{\small $i^-$} ;
\draw[black!50!white] (0,0) --(1.5,1.5);
\draw[blue!50!white]    (0,2) to[out=0,in=130] (1.5,1.5);
\draw[red!50!white]    (2,0) to[out=90,in=-50] (1.5,1.5);
\draw[blue!50!white]    (0,1.5) -- (1.5,1.5);
\draw[red!50!white]    (1.5,0) -- (1.5,1.5);
\draw[blue!50!white]    (0,1) to[out=0,in=-140] (1.5,1.5);
\draw[red!50!white]    (1,0) to[out=90,in=-130] (1.5,1.5);
\draw[blue!50!white]    (0,.5) to[out=5,in=-140] (1.5,1.5);
\draw[red!50!white]    (.5,0) to[out=85,in=230] (1.5,1.5);
\draw[red!50!white]    (2.5,0)  to[out=95,in=-50] (1.5,1.5);
\draw[blue!50!white]    (0,2.5) to[out=-5,in=140] (1.5,1.5);
\end{tikzpicture}}
\vspace{-2mm}
\caption{A hyperbolic foliation of Minkowski space.}
\label{dBS}
\end{figure}
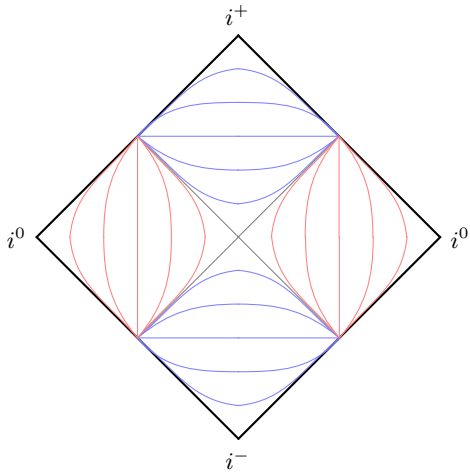

The appearance of Virasoro symmetry and soft modes in 4D gravity is reminiscent of the asymptotic  structure of 3D anti-de Sitter gravity. This relationship was anticipated in early work by de Boer and Solodukhin \cite{deBoer:2003vf}, based on the fact that 4D Minkowski space-time admits a hyperbolic foliation, as indicated in figure \ref{dBS}. The leaves of constant $X^2$ are Euclidean AdS$_3$ geometries inside the future and past lightcones of the origin (blue), and Lorentzian dS$_3$ geometries outside this lightcone (red). Lorentz transformations act as isometries on (A)dS${}_3$ and thus preserve the foliation. The celestial spheres are the common asymptotic boundaries connecting the two types of slices. Given that (A)dS${}_3$ gravity possesses an asymptotic symmetry group isomorphic to the product of two Virasoro algebras, it is natural that the asymptotic symmetries and soft dynamics of 4D gravity near the celestial boundary includes this same structure as a subsector~\cite{Cheung:2016iub}. In this section, we will use this observation to elucidate the link between celestial backreaction, symmetries and the central charge of the CCFT conformal algebra.

\subsection{Superrotated space-time}

The geometric origin of the Virasoro algebra can be understood directly from a 4D perspective by considering the vacuum solutions of the 4D Einstein equations obtained by acting with a finite superrotation on the standard Minkowski metric.  Here we will focus on the structure of pure-superrotation vacua, setting $\mathcal{C}=0$ in equation \eqref{cdeco}. Let us now try to understand what this sector looks like from the bulk. We first write the superrotated metric in Bondi gauge and then apply the coordinate transformation to cast it in the hyperbolic foliation form.

Finite superrotated vacuum solutions were studied in~\cite{Compere:2016jwb,Adjei:2019tuj}. In Bondi gauge the metric for a pure superrotation takes the form
\bea\label{met}
g_{uu}\! \is \! -1-\frac{2uV}{\sqrt{r^2+u^2V}},\qquad 
g_{ur}=\frac{-r}{\sqrt{r^2+u^2V}},\qquad
g_{uA}=-D_A\sqrt{r^2+u^2V}\\[2mm]
 g_{AB}\! \is \! (r^2+2u^2V)\gamma_{AB}+u\sqrt{r^2+u^2V}\Theta_{AB}, \qquad\qquad V=\frac{1}{8}\gamma^{AB}\gamma^{CD}\Theta_{AC}\Theta_{BD} \notag
\eea
in terms of our $\Theta$ mode above.  The above class of space-time metrics  are all related via diffeomorphisms and therefore classically equivalent. In the quantum theory, however, they become physically distinguishable. A given perturbative vacuum of the 4D bulk QFT looks like a local vacuum state only in a specific local coordinate system. Once we have specified a QFT vacuum state, local diffeomorphism invariance is spontaneously broken. It is this sense in which on can think of the variable $Z(z)$ in~\eqref{thetaf} as the Goldstone mode due to the spontaneous breaking of diffeomorphism invariance by the QFT vacuum (see also~\cite{Choi:2019rlz,Nguyen:2020hot,Pasterski:2021dqe,Pasterski:2021fjn}).

To compare to the above AdS$_3$/CFT$_2$ intuition for gravitational realizations of Virasoro symmetries, it is worthwhile to look at this metric in terms of a hyperbolic foliation of Minkowski space~\cite{deBoer:2003vf} illustrated in figure~\ref{dBS}.   For ease of presentation, we will focus on the two regions adjacent to the future celestial sphere. We will label these two regions by means of their respective points at infinity $ \textcolor{blue}{i^+} $ and $ \textcolor{red}{i^0} $. The flat Minkowski space metric in the (future) blue and (spacelike) red regions reads
\bea\label{v2}
\qquad ds^2\! \is \! -d\eta^2+\eta^2\bigl(d\rho^2+ \sinh^2\! \rho\,\spc  2\gamma_{z\bz} dz d\bz\bigr)\qquad \quad \textcolor{blue}{\ i^+} \\[2mm]
\qquad ds^2\! \is \! d\tilde{\rho}^2+\tilde{\rho}^2\bigl(-d\tilde{\eta}^2+ \cosh^2\!\tilde{\eta}\,  \spc 2\gamma_{z\bz} dz d\bz\bigr)\qquad \quad \textcolor{red}{ \ i^0}.
\eea
 To write the Bondi gauge superrotation vacuum \eqref{met} in the above foliation-friendly coordinates, we make the following coordinate identification 
\bea
\qquad  u=\eta\spc e^{-\rho},~&& ~\textstyle r=\eta \sqrt{\sinh^2\! \rho {\tiny\strut}-e^{-2\rho}V}, \qquad \quad  \textcolor{blue}{ \ i^+}\\[1mm]
\qquad u=-\tilde{\rho} \spc e^{-\tilde{\eta}},~&& ~\textstyle r=\tilde\rho \sqrt{\cosh^2\!\tilde\eta {\tiny\strut}-e^{-2\tilde{\eta}}V},\qquad \quad \textcolor{red}{ \ i^0}.
\eea
The superrotated metric~\eqref{met} in the new coordinates takes the form of a hyperbolic foliation
\bea\label{v3}
\ ds^2\!\! \is \!\! -d\eta^2\!+\eta^2\Bigl(d\rho^2\!+\left(\sinh^2\! \rho+e^{-2\rho}V\right)2\gamma_{z\bz}dzd\bz  
\spc + \spc (1\nspc -\nspc e^{-2\rho}) \bigl(\Theta^+_{zz}dz^2\! +\Theta^+_{\bz\bz}d\bz^2\bigr) \Bigr)\qquad  \textcolor{blue}{ \ i^+}\ \ \\[3mm]
\label{v4}
\  ds^2\!\! \is \!\! d\tilde{\rho}^2\!+\tilde{\rho}^2\Bigl(-d\tilde{\eta}^2 \!+\left(\cosh^2\! \tilde{\eta}+e^{-2\tilde{\eta}}V\right)2\gamma_{z\bz}dzd\bz  
\spc - \spc (1\nspc +\nspc e^{-2\tilde{\eta}}) \bigl(\Theta^-_{zz}dz^2\!+\Theta^-_{\bz\bz}d\bz^2\bigr) \Bigr)\qquad \textcolor{red}{\ i^0}. \ \
\eea
We recognize the metric on the three-dimensional leaves as the Ba\~nados geometry describing locally (A)dS space-times~\cite{Banados:1998gg}. Here we have given the $\Theta$ mode a $\pm$ superscript to indicate its restriction to the interior region to the future of the celestial sphere $\mathbb{S}^2$ or the exterior region to the past of $\mathbb{S}^2$. Requiring continuity of the metric across the future lightcone would imply that the two Goldstone modes must be identical 
\bea
\label{vacuum}
\Theta^+_{zz} = \Theta^-_{zz}. 
\eea 
The global metric obtained by combining the red and blue regions then describes a superrotated vacuum space-time with vanishing stress energy.

\subsection{Extended Virasoro algebra}

\vspace{-1mm}

If we were doing AdS$_3$/CFT$_2$, we would normally equate the $\Theta^\pm_{zz}$ in the Ba\~nados metric with the CFT stress tensor. The asymptotic Virasoro symmetry algebra in this case has non-zero central charge proportional to the AdS${}_3$ curvature radius in units of the 3D Planck length.  In our setting, the asymptotic (A)dS${}_3$ space-time has infinite radius. If instead we introduce an IR cut-off by setting the (A)dS radius equal to $R= 2/\epsilon$, measured in 3D Planck units,
applying the (A)dS/CFT dictionary to both the blue and red regions would identify two stress energy tensors whose Laurent modes 
\bea
\quad T^\pm_{zz} \! = \! \pm  \frac{i}{2\epsilon} \Theta_{zz}^\pm, \qquad \qquad  L^\pm_n=\oint\! \frac{dz}{2\pi i}\spc z^{n+1} T_{zz}^{\pm}, 
\eea
generate two Virasoro algebras with divergent central charge~\cite{Cheung:2016iub}
\bea
\label{twovirs}
[L_n^\pm,L_m^\pm]\! \is\! (n-m)L^\pm_{n+m}\mp\frac{i}{4\epsilon}(m^3-m)\delta_{n+m}.
\eea
Note that the central charge and normalization of $T^\pm$ are imaginary because of the non-standard signature and curvature radius of the hyperbolic slicing: the space-like AdS-slices in the blue interior region have a time-like radius, whereas the time-like dS-slices in the exterior region have positive curvature.  The vacuum condition \eqref{vacuum} amounts to the condition $T_n = L_n^+ + L_n^-=0$. Note that the operator $T_n$ generates a Virasoro algebra \eqref{virone} with zero central charge and setting $T_n=0$ is therefore a self-consistent first class constraint.

The above reasoning based on AdS/CFT intuition looks somewhat heuristic and has some tension with the standard celestial holographic dictionary equating the CFT stress tensor with the sub-leading superrotation memory mode~\eqref{qs}. We can resolve this tension as follows. First we observe that the combined algebra~\eqref{virtwo} generated by the $T_n$ and $\Theta_m$ derived in the previous section can be reached from an In\"on\"u-Wigner contraction of a pair of Virasoro algebras \eqref{twovirs} with divergent central charge via the identification\footnote{Note that $\Theta$ is a boundary value of the news, while $T$ involves $\frac{1}{G}$ times a $du u$ integral of the news. So $\epsilon$ is indeed dimensionless, since $[G]=L^{2}$ in 4d.} 
\bea \label{lpm}
L_n^\pm\!\! \is \!\frac 1 2 \bigl( T_n \pm \frac{i}{2\epsilon}\spc \Theta_n\bigr) , 
\eea
upon adding in the commutation relation $
[\Theta_n,\Theta_m]=-4\epsilon^2(n-m)T_{n+m}$
which vanishes in the $\epsilon \rightarrow0$ limit. This observation suggests that, starting from the two Virasoro symmetry algebras associated with the asymptotic symmetries of the  (A)dS${}_3$ leaves of the hyperbolic foliation, we can extract the modes $T_n$ and $\Theta_n$ of the CCFT stress tensor $T_{zz}^{\rm CFT}$ and Goldstone mode $\Theta_{zz}$ via the identification
\bea
T_n \! \is \! L_n^+ + L_n^- , \qquad \qquad \Theta_n \, = \, -2i\epsilon(L_n^+ - L_n^-).
\eea

Let us summarize our proposal. Up to now, it has been open question whether the celestial Virasoro algebra should  include a central charge term or not and, if so, what the value of the central charge should be. The above identifications provide a possible answer: the CCFT in fact incorporates two stress tensors, obtained by combining the stress tensor $T^{\rm \spc CFT}_{zz}(z)$ obtained from the sub-leading soft graviton mode with the superrotation Goldstone mode $\Theta_{zz}(z)$ via
\bea
\label{twotees}
T^{\smpc \pm}_{zz}(z)\!\! \is \!\! \frac 1 2 \bigl( T^{\rm \spc CFT}_{zz}(z) \pm \frac i {2\epsilon} \spc \Theta_{zz}(z)\bigr).
\eea
The Laurent modes of these two stress tensors satisfy the Virasoro algebras \eqref{twovirs} with divergent imaginary central charge. The appearance of imaginary central charge is consistent with the Hermiticity properties of the mode operators acting on the CCFT Hilbert space.

\subsection{Incorporating backreaction}\label{sec:back}

\vspace{-1mm}

In this section we will argue that examining a sector of CCFT operators that commutes with one of the subalgebras~\eqref{lpm} amounts to incorporating backreaction effects where the celestial operators excite the superrotation Goldstone mode. If we look at operators in a sector that commutes with $L_n^-$ then we are able to apply techniques associated to large central charge limits of CFT correlators. As we will now explain, this means that we will only consider operators that carry energy relative to one of the 2D stress tensors, and use the coordinate freedom to set the other stress tensor equal to zero.

The two CCFT stress tensors \eqref{twotees} can be thought of as the generators of two independent superrotation symmetry groups with infinitesimal parameters $Y^\pm$.
The stress tensors themselves transform inhomogeneously under their respective superrotations
\bea
\label{tpmtrafo}
\delta_{Y^\pm} T^\pm_{zz}\!\! \is [Q_S(Y^\pm),T^\pm_{zz}]\, =\, Y^\pm \p_z T^\pm_{zz}+ 2 \p_z Y^\pm T^\pm_{zz} \mp \frac{i}{4\epsilon} \p^3_zY^\pm.
\eea
Hence, instead of viewing $T_{zz}^\pm$ as two separate stress energy tensors, we can equally well think of them as two independent superrotation Goldstone modes. 
The infinitesimal superrotation transformation \eqref{tpmtrafo} exponentiates to the familiar inhomogeneous conformal transformation rule of a 2D CFT stress tensor involving the Schwarzian derivative.
As before, we can make the interpretation of $T^\pm_{zz}$ as superrotation Goldstone modes more manifest by writing both as a pure superrotation parametrized by two dynamical variables
\bea  
\label{zpm}
T^\pm_{zz} \!\!\is\!\!  \mp \frac{i}{4\epsilon} \{Z^\pm,z\}.
\eea 
From equation \eqref{tpmtrafo} we then see that infinitesimal superrotations generate a linear shift in $Z^\pm(z)$ via $\delta_Y Z^\pm(z) =  Y(z)$.
Hence $Z^\pm(z)$ both behave as non-linear superrotation Goldstone modes. 

Given the factorized form of the superrotation symmetry algebra, it is reasonable to assume that the CCFT operator algebra can also be naturally factorized into a tensor product of two operator algebras, spanned by primary operators ${\cal O}_+$ and ${\cal O}_-$ (and their respective descendants) that transform non-trivially under one Virasoro algebra and trivially under the other.\footnote{We use subscripts here to avoid conflating this with the $\pm$ superscript labeling in and out particles.  Unless necessary we will drop the in/out label. As we will see in the appendix, once we go to the celestial torus these can be interchanged with an appropriate $\pi$ rotation.} The leading term of the OPE of the $T^\pm_{zz}$ stress tensors of, say, a local primary operators ${\cal O}_+$ 
at the origin takes the form 
\bea
T^+_{zz}\nspc(z)\, \mathcal{O}_+\nspc(0)\! & \! \sim& \! \! \Bigl( \frac{h}{z^2}+\frac{1}{z}\p\Bigr)\mathcal{O}_+(0)\notag \\[-1.5mm]
\label{ttward}\\[-1.5mm]
T^-_{zz}\nspc(z)\, \mathcal{O}_+\nspc(0)\! & \! \sim& \! \! {\rm regular \ for}\ \  z \to 0.\notag
\eea
In the sector where $T^-_{zz} = 0$, we can identify the $Z^+(z)$ Goldstone mode introduced in \eqref{zpm} with the standard $Z(z)$ superrotation Goldstone mode introduced in \eqref{thetaf}. Hence, looking at the above form of the OPE, we deduce that in the direct neighborhood of the local operator, the $Z(z)$ behaves as follows
\bea\label{hzalpha}
-\frac{i}{2\epsilon} \{Z,z\} \! \is \! \frac{h}{z^2}  \qquad \Leftrightarrow \qquad Z(z) \, = \, z^{1-2i\epsilon h},
\eea
where we used that $\epsilon$ is small. The above two equations tell us that the insertion of a local operator $\mathcal{O}_h^+\nspc(0)$ leads to a gravitational backreaction  in the form of a superrotation that creates a conical singularity at the local operator,  with infinitesimal deficit angle equal to  $2i \epsilon h.$ This map has a branch cut connecting the north to south poles of the celestial sphere, which extends into the bulk along the locus $X^1=X^2=0$.  If we take the angular coordinate around this locus to span the usual range $\phi\in(0,2\pi]$ this locus becomes a cosmic string. 
The case where $Z(z)= z^\alpha$ with $\alpha\in\mathbb{R}$ was considered in~\cite{Strominger:2016wns}, where $\alpha = \bar{\alpha} = 1-4G\mu$ for a string of energy density $\mu$. The spinning analog is the cosmon~\cite{Deser}. Meanwhile if we instead take $\phi\in(0,\frac{2\pi}{\alpha}]$ we obtain a bulk geometry that is free of conical defects, at the price of introducing an angular deficit on the celestial sphere itself~\cite{Adjei:2019tuj}.   We will adopt this latter interpretation in what follows.

In this sector, the Goldstone and memory modes are linearly related. This linear relation implies that we can express the celestial backreaction as an operator product relation
\bea
 \Theta_{zz}(z) \mathcal{O}_+\nspc(0)\! & \! \sim &  -2{i\epsilon}\Bigl( \frac{h}{z^2}+\frac{1}{z}\p\Bigr)  \mathcal{O}_+\nspc(0)
\eea 
 between the $\Theta_{zz}$ Goldstone mode and the local operator. So in terms of the $\Theta_{zz}$ field, the backreaction is of order $\epsilon$.  We note further that, since we want the bulk geometry to be a smooth saddle, we are introducing an infinitesimal deficit angle on the celestial sphere determined by the weight of the operator under $T^{\rm CFT}$.

\section{Rindler Time and the Celestial Torus}\label{sec:ccftrindler}
\vspace{-1mm}

Let us now proceed to examine the Hilbert space description of CCFT on the celestial sphere and its analytic continuation to the celestial torus. Our guiding assumptions will be that this Hilbert space description exists, and secondly, that the Hilbert states and inner product can, via a suitable holographic dictionary, be identified with the Hilbert states and inner product of a unitary 4D quantum gravity theory. As we will argue, this 4D theory should be viewed from the perspective of a Rindler observer.

\subsection{Radial evolution on the celestial sphere}\label{sec:rad}
\vspace{-1mm}

An intrinsic way to introduce a Hilbert space description of celestial CFT is via standard radial quantization: choose an origin $(z,\bz) = (0,0)$ on the celestial sphere $\mathbb{S}^2$ and identify time translations with scale transformations $(z,\bar{z}) \mapsto (\alpha z, \alpha \bar{z})$. Finite radial time translations are generated by the evolution operator
\bea
U(t) \! \is \! e^{t (L_0 + \bar{L}_0)}.
\eea
The radial time evolution acts on local operators via
\bea\label{UO}
U(t)\mathcal{O}_\Delta (z,\bz)U^{\dag}(t)\is e^{t \Delta}\mathcal{O}_{\Delta}(e^{t} z,e^{t}\bz).
\eea
We would like to identify this transformation with unitary time evolution in the 4D quantum gravity theory. Hence, unlike in standard Euclidean 2D CFT, $U(t)$ should be a unitary operator. 
Indeed, as shown in~\cite{Pasterski:2017kqt}, a difference between celestial CFT and ordinary 2D CFT is that finite energy scattering states in the 4D theory are mapped to CCFT primary fields that carry conformal dimensions on the principal series with complex scale dimension $\Delta= 1+i\lambda$, $\lambda\in \mathbb{R}.$
This indicates that the Hilbert space inner product in CCFT (at least when restricted to the subspace of primary states) should be chosen such that $L_0+ \bar{L}_0$ is anti-Hermitian rather than Hermitian. This difference with the standard BPZ inner-product is dictated by our postulate that the 2D and 4D Hilbert space should be identified and will play a crucial role in what follows.\footnote{See~\cite{Crawley:2021ivb} for a recent effort to realize the 2D BPZ inner product from a 4D construction.}

Denoting
${\mathcal O}_\Delta (z,\bz)$ by ${\mathcal O}(\lambda, z,\bz)$, we  introduce local operators ${\mathcal O}(\tau,z,\bz)$ that act at some given instant $\tau$, via  
\bea\label{Otau}
{\mathcal O}(\tau,z,\bz) \!\is \! \int\! \frac{d\lambda \, e^{i\lambda \tau}}{2\sinh\pi \lambda} \, {\mathcal O}(\lambda,z,\bz).
\eea
The reason for including the spectral factor $(2\sinh\pi \lambda)^{-1}$ will become clear below.
The radial time evolution simultaneously shifts $\tau$ and dilates $z$ and $\bz$
\bea
U(t) {\mathcal O}(\tau, z, \bz)U^\dag(t)\is {\mathcal O}(\tau + t, e^{t} z, e^{t} \bz).
\eea
Defining rescaled celestial coordinates via 
\bea
(w,\bw) \!\is\! (e^{-\tau} z, e^{-\tau}\bz),
\eea 
this equation takes the suggestive form
\bea\label{ow}
U(t) {\mathcal O}(\tau, w, \bw)U^\dag(t)\is {\mathcal O}(\tau + t, w, \bw).
\eea
Hence we can read $U(t)$ either as the radial evolution operator in the $z$ coordinates or as the evolution operator that implements the time evolution in $\tau$. From now on we will often suppress the $w$-dependence of the local operators ${\mathcal O}(\tau, w, \bw)$ and  simply denote them as ${\mathcal O}(\tau)$.

\subsubsection{Periodicity of celestial correlators}  
\vspace{-1mm}

One can show that CCFT correlation functions of local operators ${\mathcal O}(\tau)$ in~\eqref{Otau} can be expressed in terms of the original scattering amplitudes via
\bea
\Bigl\langle {\mathcal O}^{\pm}_1(\tau_1)\ldots {\mathcal O}^{\pm}_n(\tau_n)\Bigr\rangle \is \Bigl[\; \prod\limits_i \int_0^\infty\!\! d\omega_i  \, e^{\mp i\omega_ie^{\pm \tau_i}} \Bigr] A(\omega_i)
\eea
where we've restored the $\pm$ label distinguishing the in- and outgoing asymptotic states. The above relation shows that the $\tau$ coordinate has the same properties as a Rindler coordinate: $\tau$ covers only the $u<0$ half of the light-cone time along future null infinity and the correlation functions exhibit $2\pi$ periodicity along the imaginary $\tau$ direction 
\bea
\label{kms}
\Bigl\langle {\mathcal O}^\pm_1(\tau_1) \, %
... \, {\mathcal O}^\pm_{n-1}(\tau_{n-1}
) {\mathcal O}^{\pm}_n(\tau_n)\Bigr\rangle \is 
\Bigl\langle {\mathcal O}^\pm_n(\tau_n+\! 2\pi i)\, {\mathcal O}^\pm_1(\tau_1) \, ... \, {\mathcal O}^\pm_{n-1}(\tau_{n-1})\Bigr\rangle.
\eea
Hence the correlation functions of the ${\mathcal O}(\tau)$ operators look like thermal expectation values at finite temperature $T=1/2\pi$. In the following, we will identify $\tau$ as the time coordinate of a uniformly accelerating observer moving towards the north pole of the celestial sphere.

The conclusion that celestial correlators behave as thermal expectation values looks a bit surprising, given that radial quantization in standard 2D CFT on the Euclidean  sphere produces vacuum expectation values at zero temperature. However, as emphasized above, the radial rescaling $(z,\bz)\to (e^t z, e^t \bz)$ represents unitary real time evolution in CCFT, in spite of the fact that the celestial sphere itself is Euclidean. Our postulate that the 4D Hilbert space and 2D Hilbert space should be identified dictates reality conditions that are at odds with the topology and signature of the celestial sphere.

A more appropriate setting for our purpose is to define 4D scattering amplitudes through analytic continuation from 4D Klein space $\mathbb{K}^{2,2}$ with (2,2) signature and replace the celestial sphere $\mathbb{S}^{2}$ by the celestial torus  $\mathbb{T}^{1,1}$~\cite{Atanasov:2021oyu}. 
As we will see below, a practical way to view the relationship between the two celestial spaces is to perform a Wick rotation from the celestial sphere starting from the exponential map~\eqref{expmap} taking us to the celestial cylinder illustrated in figure~\ref{fig:tauphi}.

\medskip

\subsection{Klein space and the celestial torus} Scattering amplitudes are often defined by analytic continuation from Euclidean signature.  In gravity, however, the condition of bulk diffeomorphism invariance restricts momenta to remain on-shell. For this reason, it has proven to be effective to define amplitudes in quantum gravity through analytic continuation from (2,2) signature. 
Null infinity of Klein space $\mathbb{K}^{2,2}$ takes the form of a 2-torus with (1,1) signature.
This is most easily seen by introducing the double polar coordinate parametrization of
$\mathbb{K}^{2,2}$
\bea 
(X_0, X_1, X_2,X_3) \! \is \! (q\cos \psi, \sigma\cos\phi, \sigma\sin\phi, q\sin\psi)
\eea 
 as in~\cite{Atanasov:2021oyu}. The (2,2) metric in polar coordinates reads
\bea
\label{twotworindler}
ds^2\!\! \is\!\!  - dX_0^2 + dX_1^2 + dX_2^2 - dX^2_3  =
 -dq^2-q^2 d\psi^2+d\sigma^2+\sigma^2d\phi^2.
\eea 
Future null infinity ${\cal I}$ is defined by following the light-like trajectory $q=\sigma$ to $\infty$.  It takes the form of a two torus ${\mathbb T}^{1,1}$ with metric
\bea
\label{oneonetorus}
ds^2 \! \is \! -d\psi^2 + d\phi^2 \qquad \qquad \psi \sim \psi + 2\pi, \ \phi \sim \phi + 2\pi.
\eea

A direct 4D way to go from (1,3) Minkowski space to (2,2) Klein space is to Wick rotate the $X^3$ coordinate. Alternatively, if we are only interested in asymptotic holographic data, we can directly perform the Wick rotation from the celestial sphere to the celestial torus. To do this, we suitably complexify the coordinates $(z,\bz)$ on $\mathbb{S}^2$ and write them as 
\bea
\label{complexz}
\qquad z=e^{-\tau + i \phi}, \ & & \ \bz =  e^{-\tau-i\phi} \qquad {\rm with} \qquad \tau = t - i \psi
\eea
with $t, \psi$ and $\phi$ all real. 
By the above argumentation, $\psi$ should be identified with the Euclidean Rindler time.

The 2+2 Lorentz group $SO(2,2)$ factorizes (up to a $\mathbb{Z}_2$ identification) into the product SL$(2,\mathbb{R})\times SL(2,\mathbb{R})$ of two 1D conformal groups acting on the celestial torus via M\"obius transformations
\bea
\tan\nspc 
\frac {y_\pm} 2 \to \frac{a \tan\nspc \frac{y_\pm\!} 2 + b}{c \tan \nspc \frac{y_\pm}2 + d}, \qquad y_\pm\nspc = 
\spc \psi\pm \phi. 
\eea
The lightcone coordinates $y_\pm$ are also periodic with period $2\pi$. Note, however, that the torus defined by the periods \eqref{twotworindler} is a $\mathbb{Z}_2$ orbifold of the light-like torus $\tilde{T}^{1,1}$ defined by the $2\pi$ lightlike shifts in $y^\pm$.
    
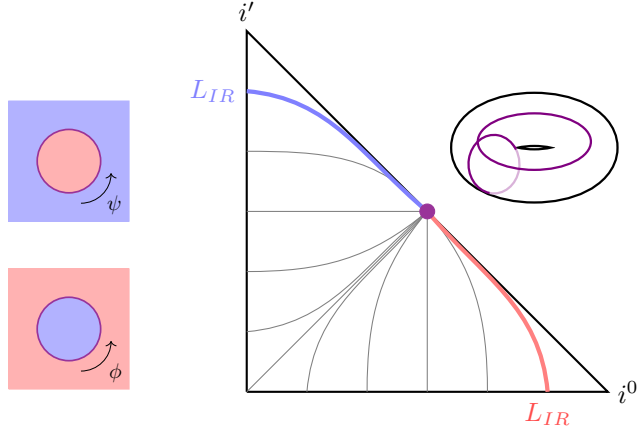
\begin{figure}[t]
\centering
\vspace{-1em}\hspace{5em}
\begin{tikzpicture}[scale=1.6]
\draw[thick] (0,0) --(3,0) node[right]{$i^0$}-- (0,3) node[above]{$i'$} --(0,0);
\draw[black!50!white] (0,0) --(1.5,1.5);
\draw[black!50!white]    (0,2) to[out=0,in=130] (1.5,1.5);
\draw[black!50!white]    (2,0) to[out=90,in=-50] (1.5,1.5);
\draw[black!50!white]    (0,1.5) -- (1.5,1.5);
\draw[black!50!white]    (1.5,0) -- (1.5,1.5);
\draw[black!50!white]    (0,1) to[out=0,in=-140] (1.5,1.5);
\draw[black!50!white]    (1,0) to[out=90,in=-130] (1.5,1.5);
\draw[black!50!white]    (0,.5) to[out=5,in=-140] (1.5,1.5);
\draw[black!50!white]    (.5,0) to[out=85,in=230] (1.5,1.5);
\draw[ultra thick,red!50!white]    (2.5,0) node[below,red!70!white]{$L_{IR}$} to[out=95,in=-50] (1.5,1.5);
\draw[ultra thick,blue!50!white]    (0,2.5) node[left,blue!50!white]{$L_{IR}$} to[out=-5,in=140] (1.5,1.5);
\filldraw[violet!80!white] (1.5,1.5) circle (1.75pt);
\end{tikzpicture}
\begin{tikzpicture}[transform canvas={scale=1.07,xscale=1,xshift=-10em,yshift=3em}]
\draw[thick,yscale=.5, double distance=1.8em] (2,5) circle (2em);
\draw[violet!30!white, thick,yscale=1.15] ($(1.6,2.57)+({.9*cos(-100)},{.9*sin(-100)})$) arc (-100:45:.9em);
\draw[violet, thick,yscale=1.15] ($(1.6,2.57)+({.9*cos(-100)},{.9*sin(-100)})$) arc (-100:-330:.9em);
\draw[violet,thick,yscale=.5] (2,5.15) circle (2em);
\end{tikzpicture}\\\hspace{6em}
\begin{tikzpicture}[transform canvas={scale=.8,xshift=-20em, yshift=12em}]
\filldraw[blue!30!white] (0,0)--(2,0) node[black,above left]{$\psi$}--(2,2)--(0,2)--(0,0);
\filldraw[red!30!white] (1,1) circle (1.5em);
\draw[violet!80!white,thick] (1,1) circle (1.5em);
\draw[->] (1.2,.3) arc (-90:0:.5);
\end{tikzpicture}\\\hspace{6em}
\begin{tikzpicture}[transform canvas={scale=.8,xshift=-20em, yshift=6em}]
\filldraw[red!30!white] (0,0)--(2,0)node[black,above left]{$\phi$}--(2,2)--(0,2)--(0,0);
\filldraw[blue!30!white] (1,1) circle (1.5em);
\draw[violet!80!white,thick] (1,1) circle (1.5em);
\draw[->] (1.2,.3) arc (-90:0:.5);
\end{tikzpicture}\\
\vspace{-3.5em}
\caption{Penrose diagram for $\mathbb{R}^{2,2}$.  Null infinity is resolved by two $AdS_3/\mathbb{Z}$ hyperboloids at fixed $X^2=\pm L_{IR}^2$ glued together along a celestial torus. These hyperboloids cap off different cycles of the celestial torus. }
\label{penrose}
\end{figure}

The boundary of $\mathbb{K}^{2,2}$ has the topology of a three sphere $S^3$ with the celestial torus at its equator. The past and future hemispheres are identified with spatial infinity $i^0$ and with future infinity $i'$. The conformal metric for both takes the form of an AdS${}_3$/$\mZ$, one with (2,1) and the other with (1,2) signature
\bea
ds_3^2 \! \is \!  -\cosh^2\nspc \rho\, d\psi^2 + \sinh^2 \nspc \rho \, d\phi^2 + d{\rho}^2 \qquad \psi \sim \psi+2\pi  \qquad  \ \textcolor{blue}{i'}\\[1.5mm]
ds_3^2 \! \is \! - \sinh^2 \nspc \tilde\eta \, d\tilde\psi^2 + \cosh^2 \nspc \tilde\eta\, d\tilde\phi^2 - d\tilde\eta^2 \qquad\, \tilde\phi \sim \tilde\phi+2\pi \qquad  \ \textcolor{red}{i^0}.
\eea
The identifications $\tilde \psi \sim \tilde\psi+2\pi$ and  $\phi \sim \phi+2\pi$ are automatically imposed by imposing smoothness at $\tilde\eta=0$ and $\rho=0$.
The two AdS${}_3$/$\mZ$ caps are glued together via the identification $\tilde{\phi} = \phi$ and $\tilde{\psi} \, = \, \psi$.

It is useful to compare the metric \eqref{twotworindler} with the standard Rindler metric in Lorentzian signature
\bea
\label{threeonerindler}
ds^2\! \is\! 
-\xi^2 d\tau^2 + d\xi^2+d\sigma^2+\sigma^2d\phi^2.
\eea
The metrics \eqref{twotworindler} and  \eqref{threeonerindler} can be mapped to each other through the identification between the corresponding Euclidean metrics. We can double Wick rotate the (2,2) signature metric to Euclidean signature by replacing the $q$ coordinate by $q_{\nspc E} = i q$. 
Similarly, we obtain the Euclidean version of the Rindler metric  \eqref{threeonerindler} by Wick rotating the Rindler time coordinate to $\tau_{\nspc E} = -i \tau$. Equating the two metrics
\bea
\label{fourzerorindler}
ds^2_{\!\spc{E}} \! \is\! 
dq^2_{\!\spc{E}}  + q^2_{\!\spc{E}}  \spc d\psi^2 +d\sigma^2+\sigma^2d\phi^2 
 \, = \,
d\xi^2+\xi^2 d\tau_E^2 + d\sigma^2+\sigma^2d\phi^2 
\eea
leads us to identify the Rindler space coordinate $\xi$ with the Euclidean $q_{\nspc E}$ coordinate of the Klein space-time and the celestial time coordinate $\psi$ with the Euclidean Rindler time coordinate $\tau_{\nspc E}$ 
\bea
 \xi = q_{\nspc E} = i q, \quad & & \quad \psi = \tau_{\nspc E} = i \tau
\eea
on the locus $t=0$.

\subsection{Superrotated Klein space}

\vspace{-1mm}

As described the previous section for Minkowski space-time, one can again write a general class of vacuum solutions obtain by acting the with finite superrotation transformations on Klein space. Here we only state the form of the solutions in both regions
\bea\label{tmm2}
ds^2\! \is \! -d\eta^2+\eta^2\Big(d\rho^2-\frac{1}{4}(e^{\rho} dx^++4e^{-\rho} \Theta^+_{--}dx^-) (e^{\rho} dx^-+4e^{-\rho} \Theta^+_{++}dx^+)\Big) \qquad\ \ \textcolor{blue}{i'}\\[2mm]
ds^2\! \is \! d\tilde\rho^2+\tilde\rho^2\Big(-d\tilde\eta^2-\frac{1}{4}(e^{\tilde\eta} dx^+-4e^{-\tilde\eta} \Theta^-_{--}dx^-) (e^{\tilde\eta} dx^--4e^{-\tilde\eta}\Theta^-_{++} dx^+)\Big)  \qquad \textcolor{red}{i^0}.
\eea
In the vacuum solutions without any matter stress energy in the bulk, we need to impose the condition that $\Theta^\pm = \Theta^\pm$.
The standard Klein vacuum solution corresponds to special case that $\Theta^\pm_{--} = \Theta^\pm_{++} = 1/4$. The holographic CFT interpretation of this non-zero value of $\Theta^\pm$ is that the CCFT on the celestial torus has both a finite Casimir energy and a finite temperature. Indeed, as before, we can make a holographic dictionary between the asymptotic symmetry groups and Goldstone variables $\Theta^\pm$ of the (A)dS${}_3$  at $i'$ and $i^0$ and the appearance of two stress energy tensors $T^\pm$ in the dual CCFT on $\mathbb{T}^{1,1}$. We will now describe some general properties of the partition function and correlation functions of this CCFT.

\def\TFD{{\mbox{TFD}}}

\medskip

\section{CFT on the Celestial Torus}\label{CCFTorus}
\vspace{-1mm}

Via a slight generalization of the dictionary outlined above for the celestial sphere (which we review in appendix~\ref{sec:TOp}), one can identify scattering amplitudes in (2,2) signature space-time with correlation functions of a putative 2D CFT defined on ${\mathbb T}^{1,1}$. It is reasonable to assume that this CCFT is identical to the one obtained by Wick rotating the Euclidean CCFT defined on the celestial sphere and placing it on $\mathbb{T}^{1,1}$. 

We have shown that the chiral symmetry algebra of this CCFT contains two Virasoro algebras $L_n^+$ and $L_n^-$ with imaginary central charge $c\to \pm i\infty$ and
that this result is naturally linked to the way in which the celestial geometry is embedded in asymptotic infinity. In particular, in (2,2) signature, the celestial torus forms the interface between two halves of the asymptotic three sphere $S^3$, each of which take the form of an AdS${}_3/\mathbb{Z}$ space-time with infinite curvature radius. Hence, if we would cut the asymptotic three sphere open at the equator, along the celestial torus, we create two disconnected AdS${}_3$ space-times with asymptotic boundaries. It is natural to interpret the two Virasoro algebras as the edge modes associated with the asymptotic symmetry groups of theses two  AdS${}_3$  hemispheres. Since the two AdS${}_3$ space-times have infinite radius, the corresponding Virasoro algebras have infinite central charge. The central charge is imaginary because both hemispheres have a non-standard signature.

 In the following, we will assume that the Hilbert space and  operator algebra of the CCFT can be factorized into a tensor product of two Hilbert spaces and two operator algebras, spanned by highest weight states and primary operators ${\cal O}^+$ and ${\cal O}^-$ and their respective $L_{-n}^+$ and $L_{-n}^-$ Virasoro descendants. To simplify the discussion and notation, we  will focus our discussion below on the partition function and correlation functions of operators restricted to one of these two sectors.   We will not explicitly indicate the $+$ or $-$ label.
 
\def\NN{\mbox{
$N$}}

\def\NM{\mbox{
$M$}}

\addtolength\parskip{1mm}

\def\uU{\mbox{\small$U$}}
\def\vV{\mbox{\small$V$}}
\def\cZ{{\cal Z}}
\def\ss{i}
\def\ttau{\mbox{$\sigma$}}

\subsection{Partition function}\label{sec:spec}

\vspace{-1mm}

The first step in understanding the CCFT on the celestial torus is to write the partition function, defined by performing the CCFT path-integral on $\mathbb{T}^{1,1}$.  First, let us determine the modular parameter.  Noting that the Lorentzian torus is obtained by analytic continuation from the  Euclidean torus, or equivalently, by taking a real slice of the complexified torus $ds^2 = (dx + \ttau dy) (d \bar{x} + \bar{\ttau} d\bar{y})$ with modular parameter $\ttau$ via  the identification $
x = \bar{x} = \phi$, $y = \bar{y} = \psi$ and $\ttau = -\bar\ttau= 1.$
The sum over states in the partition function are therefore weighted by $U = q^{L_0} \bar{q}^{\bar{L}_0}$ with $
q = e^{2\pi i \sigma}   =  e^{2\pi i}$ and $\bar{q} = e^{-2\pi i \bar\sigma}  =  e^{2\pi i }.$ So the partition function on the celestial torus takes the form
\bea\label{Z}
\cZ \is \tr\bigl(
e^{2\pi i (L_0+ \bar{L}_0)} \bigr). 
\eea
Here the trace is defined over the Hilbert space of all primary and descendant states in the CCFT, as defined via the standard operator state correspondence.

Concretely, we again wish to define the CCFT Hilbert space on the celestial torus via an operator state correspondence of the form $
|h,\bar{h}\rangle = \mathcal{O}_{{\Delta},{J}}(0,0)|0\rangle$, where $(h, \bar{h}) = \bigl(\frac 12 (\Delta\! +\nspc J),\frac 12 (\Delta\! -\nspc J)\bigr)$. 
Note, however, that the states contributing to the trace in the $\mathbb{T}^{1,1}$ partition function are not created by local operators on the celestial torus itself:  the point $(z,z) = (0,0)$ does not describe a point on ${\mathbb T}^{1,1}$. Instead, we will identify it with the north pole of the celestial sphere.

\begin{figure}[t]
\begin{center}
\raisebox{1.8cm}{$e^{2\pi i(L_0+\bar{L}_0)} \ = \ $}~~~\raisebox{5mm}{\begin{tikzpicture}[scale=.69]
\draw[yscale=.35] (1.57,1.15) circle (1.61em);
 \draw[black,dashed,fill=white,yscale=.4] (1,1) to[out=90,in=180] (1.57,1.65) to[out=0,in=90] (2.15,1);
\draw[black,yscale=.4] (1.57,-.21) circle (1.62em);
\draw[thick] (1,-0.1) arc (342:16:.85);
\draw[thick] (2.15,-.12) arc (352:8:1.95);
\end{tikzpicture}}~~~~\raisebox{1.75cm}{$= \ {{\raisebox{-6pt}{\large $\sum$}} \atop{\raisebox{-6pt}{\scriptsize${i,N}$}}}\ e^{-2\pi \lambda_i}$}
\raisebox{3mm}{\begin{tikzpicture}[scale=.395]
\tikzmath{\x1 = 5; \y1 =1; }
  \draw[black,thick,fill=white] (-3.75+\x1,-1.25) to[out=90,in=180] (-2+\x1,.25) to[out=0,in=90] (-.25+\x1,-1.25);
\draw[dashed,xscale=.7,yscale=.725] (3.3,-5.35) arc (320:40:5);
\filldraw[black, thick] (-2+\x1,-3-\y1) circle (.2em) node[below,fill=white]{$\raisebox{-1pt}{\footnotesize $\ \, |\ss,N\rangle_{{\!}_{ R}}$}$};
 \draw[black,thick] (-3.75+\x1,-1.25-\y1) to[out=-90,in=180] (-2+\x1,-3-\y1);
  \draw[black,thick]   (-2+\x1,-3-\y1) to[out=0,in=-90] (-.25+\x1,-1.25-\y1);
\draw[black,fill=white,yscale=.25] (-2+\x1,-4*\y1-5.25) circle (5em);
\filldraw[black, thick] (-2+\x1,0.25) circle (.2em) node[above,fill=white]{$\raisebox{1pt}{\footnotesize
${}_{{\!}_{ R}\!}\langle \ss,N| $}$};
\draw[yscale=.25] (-2+\x1,-5.25) circle (5em);
 \draw[black,dashed,fill=white,yscale=.25] (-3.75+\x1,-5.65) to[out=90,in=180] (-2+\x1,-2.75) to[out=0,in=90] (-.25+\x1,-5.25);
\end{tikzpicture}}
\vspace{-7mm}
\end{center}
\caption{The evolution operator $U= e^{2\pi(L_0+ \bar{L}_0)}$ that brings states around the celestial torus can be expanded in states defined via the operator state correspondence of the CCFT on the celestial sphere. \label{t11decon}}
\end{figure}
According to the metric on $\mathbb{T}^{1,1}$, $\psi$ is a periodic Lorentzian time coordinate. Hence it looks like the  $\mathbb{T}^{1,1}$ has closed time-like curves. This seems problematic, since we would like to consider the celestial CFT as a physical theory with consistent causal dynamics. However, just as for the celestial sphere, the signature of time direction on the celestial torus is opposite to what one would have expected based on the corresponding 4D interpretation of this time flow. As we have argued, evolution in the $\psi$ direction should be viewed as Euclidean Rindler time evolution. Hence, as before, we will need to choose our Hilbert space inner product such that the operator $U = e^{2\pi i (L_0 + \bar{L}_0)}$ that implements a full $2\pi$ shift in the $\psi$ coordinate has purely real eigenvalues of the form $e^{-2\pi \lambda} <1$, that in the mapping to 4D Rindler space represent the Boltzmann weights of the thermal Minkowski vacuum. 

Based on the mapping between the 2D and 4D Hilbert space, we again deduce that the conformal weights are captured by data on the principal series, and thus take the form $\Delta = 1 + i \lambda$. Using the Laurent modes of the stress tensor, we generate Virasoro descendants 
\bea \label{descend}
|i, \NN\rangle \propto
\prod_{j,\bar{j}} L_{-n_j}  \bar{L}_{-\bar{n}_{\bar j}}
|h_i, \bar{h}_i\rangle
\eea
where $N$ is the short-hand label for the collection of all descendant states with total conformal weight $
(L_0 + \bar{L}_0)|i, \NN\rangle = (\Delta_{i}\! + \NN) |i, \NN\rangle $
with  $\NN \equiv \sum_j n_j + \sum_{\bar j} \bar{n}_{\bar j} .$
The partition function on $\mathbb{T}^{1,1}$ is a trace over the Hilbert space of all CCFT states.  Since in each case the conformal dimension of the descendent states is shifted by integer values, the unit modular parameter implies that all states in the same conformal tower appear with the same weight in the partition function~\eqref{Z}.
\bea
e^{2\pi i (L_0+ \bar{L}_0)} \is \sum_{\ss,N}\, e^{ 2\pi i  \Delta_\ss}|\ss,\NN\rangle \langle \ss,\NN| \, = \, \sum_{\ss,N}\, e^{-2 \pi\lambda_\ss} |\ss,\NN\rangle\langle \ss,\NN|.
\eea
Upon taking a trace, we find 
\bea
\label{ceezee}
{\cal Z} \is \sum_{\ss,N}\, e^{ 2\pi i  \Delta_\ss}\, = \, \sum_{\ss,N}\, e^{- 2 \pi\lambda_\ss}.
\eea
We see that the partition sum contains a formally divergent factor in the form of the unrestricted sum over descendant states.

\subsection{Goldstone modes}

\vspace{-1mm}

The partition function $\cZ$ in \eqref{ceezee} contains a divergent factor $\cZ_0$ due to the presence of an infinite tower of Virasoro descendants. Their contribution is not suppressed, since, for the specific shape of the celestial torus, all descendants of a given primary state with conformal dimension $\Delta = 1+i\lambda$ acquire the same Boltzmann weight $e^{-2\pi \lambda + 2\pi i N} =e^{-2\pi \lambda}$. The divergent factor $\cZ_0$ and infinite tower of descendants are both linked to the emergence of massless Goldstone modes associated with the Virasoro group Diff($S^1) \times$Diff($S^1)$. Indeed, $\cZ_0$ can be shown to equal to the volume of the Virasoro group. In anticipation of their relevance the Lyapunov behavior of OTOCs, let us make this Goldstone mode contribution to the partition sum more explicit. 

It is natural to define the celestial torus partition function with $q = e^{2\pi i}$ as the limit of a finite expression
\bea
\cZ\!\! \is \! \lim_{q\to e^{2\pi i }} \, 
\sum_i \, \raisebox{1pt}{$\chi$}\raisebox{-1pt}{${}_{h_i}\!$}(q)\,\raisebox{1pt}{$\chi$}\raisebox{-1pt}{${}_{\bar{h}_i}\!$}(\bar{q})\spc.
\eea 
Here we introduced the Virasoro characters $\chi_h(q)$, defined as the trace of $q^{L_0 - \frac{c}{24}}$ 
over the  Virasoro representation with highest weight $h$ and central charge $c$. As explained above, the conformal weights and the central charge are both imaginary. In particular, the asymptotic flat space-time corresponds to the $c\to i\infty$ limit. Plugging in the explicit form of the spectrum gives
\bea 
\cZ\!\! \is \! 
\sum_i  \spc e^{-2\pi \lambda_i} \cZ_{h_i\nspc, \bar{h}_i} \qquad \qquad\
\cZ_{h,\bar{h}} = \spc \lim_{q\to 1}\, \raisebox{1pt}{$\chi$}\raisebox{-1pt}{${}_{h}\!$}(q)\,\raisebox{1pt}{$\chi$}\raisebox{-1pt}{${}_{\bar{h}}\!$}(\bar{q}).
\eea 
We will now show that the prefactors $\cZ_{h,\bar{h}}$ are in fact all identical and equal to the volume of Diff(S${}^1$).

The Virasoro character with conformal weight $h$ and central charge $c$ can be represented as a path integral over a co-adjoint orbit of Diff(S$^1$), the group of diffeomorphisms of the unit circle in the complex $z$ plane~\cite{Alekseev:1988ce,Alekseev:1990mp,Cotler:2018zff,Mertens:2019tcm}
\bea
\label{ascharacter}
\chi_h( q)\!\! \is\!\! \int \nspc \left[\mathcal{D}f\right] \, e^{\raisebox{1pt}{\smpc{\footnotesize $\frac {ic} {24\pi}$\!\!\!}
\footnotesize $\int\! dt\spc dz \, ( \Omega  - 
  T )$}}
\qquad{\rm with}\ \ \ \left\{
\begin{array}{cc} 
{T\spc =\;} & \!\!\!\!\!\! -{\{ f, z\} 
\qquad \qquad \qquad}\\[2.5mm]
{\Omega  \spc =\;} & \!\!\!\!\!\!\!
{ \mbox{\large ${\frac{\dot{f}}{2 f'} \nspc\bigl (\nspc \frac{f'''}{f'}}$}\nspc 
 - \mbox{\small $2$}  \mbox{\large ${\bigl(\nspc\frac{f''}{f'}\nspc \bigr)^{\mbox{\scriptsize 2}}}\spc \bigr)$}}\\[3.85mm]
{f(e^{2\pi i} z)}\!\!\!\!\! &\; {=  e^{2\pi i \alpha}f(z). \qquad }
 \end{array}\right.
\eea
Here the functional integral runs over a two dimensional field $f(z,t)$, with $t$ a periodic Euclidean time coordinate with period $\beta_{\rm 2D} = \log q$ and $z$ restricted to the unit circle. Hence, for fixed $t$, the function $f(z,t)$ represents an element of Diff(S$^1$) and the integral runs over all paths in the Virasoro group. Diff(S$^1$) is a symplectic manifold endowed with a canonical symplectic two-form $\omega$ and \eqref{ascharacter} is a path integral over this phase space. $\Omega$ is the geometric Virasoro action, defined via the property that (upon replacing $\dot{f} dt = \delta f$, so that $\Omega$ becomes a one-form on the Virasoro group manifold) its exterior derivative with respect to $f$ equals $\delta\Omega = \omega$
with $\omega$ the canonical symplectic two-form on Diff(S$^1$). The geometric action and symplectic form are designed such that, upon quantization, the operators $T$ satisfy the Virasoro algebra
\bea
\label{virasorohbar}
[T(z_1),T(z_2)]\!\! \is \!\! - \hbar (T(z_1) + T(z_2) ) \delta'(z_{12}) + \frac{\hbar}{2}\delta'''(z_{12}), \qquad \qquad \hbar = \frac{6}{c}.
\eea
Moreover, the twisted boundary condition $f(e^{2\pi i} z) = e^{2\pi i \alpha} f(z)$ specifies the specific co-adjoint orbit associated with the highest weight representation with conformal weight $h$ related to the twist angle $\alpha$ via
\bea
\frac{24 h}{c} = 1-\alpha^2,
\eea
matching~\eqref{hzalpha} above. Note that in the $c \to i \infty$ limit with $h$ finite, the twist angle goes to zero ($\alpha\rightarrow 1$). From the point of view of the quantum theory \eqref{ascharacter} of the Virasoro algebra \eqref{virasorohbar}, sending $c \to i\infty$ corresponds to taking a classical $\hbar \to 0$ limit. In this limit, the trace over the Hilbert space reduces to an integral over the phase space.

As shown in \cite{Mertens:2017mtv}, in the scaling limit $q\to 1$, $c\to i \infty$ with $q^{c/12} = e^{-\beta}$ fixed, the  partition function \eqref{ascharacter} of the geometric Virasoro theory reduces to the partition function of Schwarzian quantum mechanics
\bea
 \lim_{\mbox{\footnotesize ${{q\to 1},{c\to i\infty}}\atop{q^{\mbox{\tiny $c/12$}}}= \mbox{\scriptsize\,\spc $e$}^{\mbox{\tiny $-1/\beta$}}$}}{\chi_h(q)}
\!\!\is\!  \int\nspc [\mathcal{D}f]~e^{\mbox{\footnotesize $\frac  1 {\beta}\int\! dz \,\{f, z\}$} } 
\eea
up to a divergent prefactor of the form $e^{S_0-\beta E_0}$. Here the functional integral runs over all diffeomorphisms $f(z)$ of the unit circle. In the  strong coupling $\beta \to \infty$ limit of the Schwarzian QM,  the overall divergent prefactor in the CCFT partition function reduces to the volume of the Virasoro group Diff(S$^1$) $\times$ Diff(S$^1$). More generally, if we relax this limit, we would find that the celestial CFT contains a soft sector described by Schwarzian quantum mechanics with coupling $\beta =- \frac{12}{c\log q}$. This suggests a possible link between celestial CFT and SYK-like dynamics \cite{PV3}.

\subsection{Correlation functions }

\vspace{-1mm}

Up to now, we have followed the standard philosophy and used the properties and symmetries of scattering amplitudes and the asymptotic geometry to extract information about the properties of the celestial CFT. In what follows, we will aim toward setting up the CCFT as a physical quantum system equipped with a Hilbert space and intrinsic dynamics. We will adopt the following guiding principles: 

\addtolength\parskip{-1mm}
\begin{enumerate}
\addtolength\parskip{-1mm}
    \item The spectrum of Hilbert states in CCFT, obtained by radial quantization  on the celestial sphere ${\mathbb S}^2$, is isomorphic to the spectrum of the bulk theory in the Rindler wedge seen by a single Rindler observer.
    \item Any observable that we can compute in the Rindler wedge has a celestial holographic CFT dual. In and out states in the wedge are created by operators localized at the north and south pole of ${\mathbb S}^2$.
    \item The thermal mixed state seen by the Rindler observer corresponds to a thermal state in CCFT on $\mathbb{S}^2$. The Minkowski vacuum maps to a TFD state entangling two CCFTs defined on two copies of~${\mathbb S}^2$.
\addtolength\parskip{1mm}
\end{enumerate}

The previous subsection examined how to cut open and insert a complete set of states on the celestial torus.  Combining this with our construction of local primary operators on $\mathbb{T}^{1,1}$ detailed in appendix~\ref{sec:TOp}, we can give a 2D description for evaluating their correlation functions.  In what follows, we will be interested in the time ordering dynamics and will suppress the $\phi$ coordinate. Unless otherwise specified the same statements will hold for operators smeared over the $\phi$ cycle.
Using our discussion of the modular parameter in the previous section, the $\tau$-ordered correlator can be written as a trace 
over a complete set of states in the Hilbert space
\bea\label{correlator}
\bigl\langle {\mathcal O}_1(\tau_1) \, ...  \, {\mathcal O}_n(\tau_n)\bigr\rangle \is {\rm Tr}\Bigl(e^{2\pi i(L_0+\bar{L}_0)} {\mathcal O}_1(\tau_1) \, ...  \,\spc {\mathcal O}_n(\tau_n)\Bigr).
\eea 
Alternatively, we can represent this correlation function as an expectation value in the thermofield double state
\bea
\bigl\langle {\mathcal O}_1(\tau_1) \, ...  \, {\mathcal O}_n(\tau_n)\bigr\rangle \is  \bigl\langle {\rm TFD} \bigl| 
 {\mathcal O}_1(\tau_1) \, ... \, {\mathcal O}_n(\tau_n) \bigr|\rm{TFD}\bigr\rangle.
\eea
The thermofield double state is an entangled state between two copies of the CCFT Hilbert space, with the property that the mixed state in one copy obtained by tracing out the other copy is given by the thermal density matrix.
\bea\label{torustfd}
\raisebox{.7cm}{$|{\rm TFD}\rangle ={{\raisebox{-6pt}{\large $\sum$}} \atop{\raisebox{-6pt}{\scriptsize ${i,N}$}}}\, e^{- \pi\lambda_\ss} |\ss,\NN\rangle_{{\!}_{ R}} |\ss,\NN\rangle_{{\!}_{ L}}  = \ {{\raisebox{-6pt}{\large $\sum$}} \atop{\raisebox{-6pt}{\scriptsize ${i,N}$}}}\ e^{-\pi \lambda_i}$}~
\begin{tikzpicture}[scale=.6]\draw[thick,yscale=.5] (0.666,0) circle (1.9em) node{$\raisebox{1cm}{\small \,$|\ss,\NN\rangle_{{\!}_{ L}}$}$};
\draw[thick,yscale=.5] (3.666,0) circle (1.9em) node{$\raisebox{1cm}{\,\small $|\ss,\NN\rangle_{{\!}_{ R}}$}$};
\draw[thick] (1.33,0) arc (-180:0:.833);
\draw[thick,yscale=.95] (0,0) arc (-180:0:2.167);
\end{tikzpicture}
\eea
We see that the Minkowski vacuum appears to prepare an entangled state on the $S^1\times S^1$.   Via our interpretation of the $\psi$ evolution as imaginary Rindler time, it is natural to identify this TFD state with the Minkowski vacuum as experienced by the Rindler observer.  
\bea
\bigl\langle {\mathcal O}_1(\tau_1) \, ...  \, {\mathcal O}_n(\tau_n)\bigr\rangle\is \sum_{\ss,N} e^{-2\pi \lambda_\ss} \bigl\langle \ss,\NN \bigr| {\mathcal O}_1(\tau_1) \, ... \spc \, {\mathcal O}_n(\tau_n) \bigl|\ss,\NN\bigr\rangle.
\eea
For the two point function we have
\def\rr{\beta}
\bea\label{complete}
\bigl\langle {\mathcal O}_2(\tau) \, {\mathcal O}_1(0)\bigr\rangle
\!\is\!\!\!\! \sum_{i,j,N,\tilde{N}}\! e^{-(2\pi\nspc-\nspc i\tau) \lambda_i \nspc-\nspc i\tau \lambda_j  } \bigl\langle i,\NN \bigr| \spc {\mathcal O}_1\spc \bigl|j,\tilde\NN\bigr\rangle\bigl\langle j,\tilde\NN\bigr|\spc {\mathcal O}_2\spc \bigl|i,\NN\bigr\rangle \ = \; \sum_{i,j} \raisebox{-.9cm}{
\begin{tikzpicture}[xscale=0.36,yscale=1, rotate = 0]
\draw[thick] (-.7,0)--(0,0) -- (.7,0) -- node[above]{$j$}
(1,0) -- (2.25,0);
\draw[thick] (-.75,-1) -- (.8,-1) node[below]{$i$}--(2.25,-1) ;
\draw[thick]  (1.85,0) -- (1.85,.5) node[right]{${\cal O}_1$} -- (1.85,.7);
\draw[thick]  (-.25,0) -- (-.25,.5) node[left]{${\cal O}_2$} -- (-.25,.7);
\draw[thick,xscale=3]    (-.2,-1) to[out=180,in=-90]  (-.7,-.5) to[out=90,in=180] (-.2,0);
\draw[thick,xscale=3]    (.75,-1) to[out=0,in=-90] (1.25,-.5) to[out=90,in=0] (.75,0);
\end{tikzpicture}}\,.\ \ \
\eea
Here on the right we have captured the sum over intermediate Virasoro sectors in graphical notation. Due to the left-right factorization of the conformal algebra, the correlation function are given by a (possibly infinite or continuous) sum of terms that factorize into a product of left- and right-moving conformal blocks.  In the following section we will use features of 2D Virasoro blocks to examine OTOCs in this sector. The conformal block decomposition of perturbative amplitudes in celestial CFT has been studied in~\cite{Lam:2017ofc,Fan:2021isc,Atanasov:2021cje}. 

\medskip

\section{Signatures of Chaos in Celestial CFT}\label{OTOCinCCFT}

\setcounter{secnumdepth}{4}
\vspace{-1mm}

Equipped with our 4D and 2D understandings of the CCFT dynamics and its relation to Rindler dynamics, we are prepared to study the out of time ordered correlation functions and identify the onset of chaotic behavior in both pictures. In this section, we will first introduce the OTOCs of interest and then outline and compare three basic methods of computation. The first two rely on standard tools from 2D conformal field theory, namely i) the known expressions for the monodromy matrices that relate different operator orderings of Virasoro conformal blocks and ii) the analytic properties of a suitably chosen vacuum block. Both methods are closely related and aided by fact that CCFT conformal blocks arise from taking the large central charge limit of Virasoro blocks. We then compare the CCFT results with the prediction obtained by including the gravitational backreaction of the 4D Einstein theory. We will see that all methods of computation will give the same answer. This match is not coincidental, but the consequence of a direct geometric correspondence between the gravitational backreaction and the monodromy properties of the CCFT conformal blocks.   

\medskip 
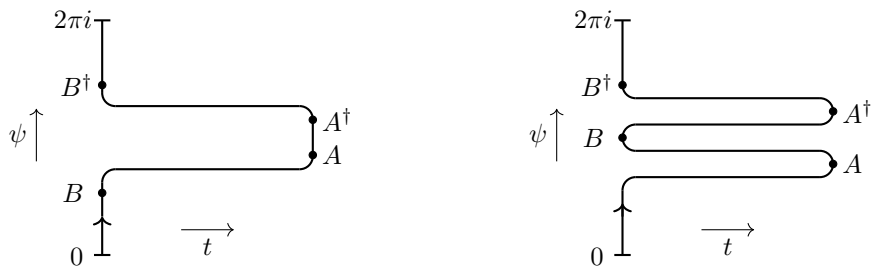
\begin{figure}[hbtp]
\centering
\vspace{-0.2em}
  \raisebox{-.75cm}{
\begin{tikzpicture}[xscale=-.7,yscale=.7]
\draw[thick,|->] (0,-1.5) node[left]{$0~$}--(0,-.75);
\draw[thick] (0,-.75)--(0,-.1);
\draw[thick,-|] (0,1.6) -- (0,3) node[left]{$2\pi i$};
\draw[thick] (-4,0.4) -- (-4,1.1);
\draw[thick] (-.25,0.15) -- (-3.75,0.15);
\draw[thick] (-3.75,1.35) -- (-.25,1.35);
\filldraw[black] (0,1.75) circle (2pt) node[left]{$B^\dagger$};
\filldraw[black] (0,-.3) circle (2pt) node[left]{$B~$};
\filldraw[black] (-4,.42) circle (2pt) node[right]{$A$};
\filldraw[black] (-4,1.09) circle (2pt) node[right]{$A^\dagger$};
\draw[thick]    (0,-.1) to[out=90,in=0]  (-.25,0.15);
\draw[thick]  (-.25,1.35) to[out=0,in=-90]    (0,1.6);
\draw[thick]  (-3.75,0.15) to[out=180,in=-90]    (-4,.4);
\draw[thick]  (-3.75,1.35) to[out=180,in=90]    (-4,1.1);
\draw[<-] (-2.5,-1)--node[below]{$t$}(-1.5,-1);
\draw[->] (1.25,-.2+.5)--node[left]{$ \psi$}(1.25,.8+.5);
\end{tikzpicture}
}~~~~~~~~~~~~~~~
  \raisebox{-.75cm}{
\begin{tikzpicture}[xscale=-.7,yscale=.7]
\draw[thick,|->] (0,-1.5) node[left]{$0~$}--(0,-.5);
\draw[thick] (0,-.75)--(0,-.25);
\draw[thick,-|] (0,1.75) -- (0,3) node[left]{$2\pi i$};
\draw[thick] (-.25,0) -- (-3.75,0);
\draw[thick] (-3.75,.5) -- (-.25,.5);
\draw[thick] (-0.25,1) -- (-3.75,1);
\draw[thick] (-3.75,1.5) -- (-.25,1.5);
\filldraw[black] (0,1.75) circle (2pt) node[left]{$B^\dagger$};
\filldraw[black] (0,.75) circle (2pt) node[left]{$B~$};
\filldraw[black] (-4,.25) circle (2pt) node[right]{$A$};
\filldraw[black] (-4,1.25) circle (2pt) node[right]{$A^\dagger$};
\draw[thick]    (0,.75) to[out=90,in=0]  (-.25,1);
\draw[thick]    (0,-.25) to[out=90,in=0]  (-.25,0);
\draw[thick]  (-.25,.5) to[out=0,in=-90]    (0,.75);
\draw[thick]  (-.25,1.5) to[out=0,in=-90]    (0,1.75);
\draw[thick]  (-3.75,0) to[out=180,in=-90]    (-4,.25)
to[out=90,in=180]    (-3.75,.5);
\draw[thick]  (-3.75,1) to[out=180,in=-90]    (-4,1.25)
to[out=90,in=180]    (-3.75,1.5);
\draw[<-] (-2.5,-1)--node[below]{$t$}(-1.5,-1);
\draw[->] (1.25,-.2+.5)--node[left]{$ \psi$}(1.25,.8+.5);
\end{tikzpicture}
}

\caption{The TOC and OTOC contours. The imaginary time direction is compact.}
\label{OTOCcontour}
\end{figure}

\vspace{-0mm}

\subsection{OTOCs on the celestial torus} 
\vspace{-1mm}

We are interested in studying the OTOCs of four local operators in celestial CFT on the celestial torus. As explained in the previous section, the Hilbert space on $\mathbb{T}^{1,1}$ is obtained by acting with local operators on the thermofield double state and, correspondingly, correlators are given by expectation values between two TFD states. For simplicity, we will only keep track of the $t$ dependence of the correlation function. 

Consider an operator $B(t_0)$ smeared in the $\phi$ direction inserted at time instance $t_0$ and an operator $A(t_1)$ smeared in the $\phi$ direction inserted at a later time $t_1$.  We will look at two types of states: the state $|x\rangle$ prepared by the time ordered configuration and
and the state $|y\rangle$ prepared by a time fold configuration
\bea
|x\rangle\! \is\!A(t_1)B(t_0)|{\TFD} \rangle\\[2mm]
|y\rangle\! \is\!B(t_0)A(t_1)|\TFD\rangle.
\eea
We can write the TO and OTO four-point correlation functions as inner products of these states
\bea
\langle x |y\rangle \! \is\! 
{\tr}\bigl(e^{2\pi i (L_0+\bar{L}_0)}B^\dagger(t_0) A^\dagger(t_1) B(t_0)A(t_1)\bigr)\\[2mm]
\langle x|x\rangle\! \is \!
\tr\bigl(e^{2\pi i (L_0+\bar{L}_0)}B^\dagger(t_0) A^\dagger(t_1) A(t_1)B(t_0)\bigr).
\eea
The overlap $\langle x |y\rangle $ between the time ordered ket-state $|y\rangle$  and out-of-time-ordered bra-state $\langle x|$ is the OTOC. The corresponding $\tau$ time-contour is indicated in figure \ref{OTOCcontour}.  The time ordered four-point function defined by the norm $\langle x |x\rangle$ diverges and needs to be regulated by point-splitting. To leading order in the point-splitting distance, it factorizes into the product of two-point functions on the sphere, times the partition function on the celestial torus. It will be useful to consider the normalized ratio of the OTOC and TOC
\bea
\label{otocratio}
\frac{\langle x|y\rangle }{\langle x|x\rangle}\! \is \! 
\frac{\langle B^\dagger A^\dagger B A \rangle }{\langle B^\dagger A^\dagger A B\rangle} \, = \, \frac{\langle B^\dagger A^\dagger B A \rangle }{\langle B^\dagger B\rangle\langle A^\dagger A \rangle}.
\eea 
Below we will describe three ways of computing this ratio.

As a first preparation, we start by writing the time-ordered and out-of-time-ordered four point functions as a sum over conformal partial waves
\bea
\label{toc}\langle B^\dagger A^\dagger    A B\rangle \! \is \!  \sum_{{ijkl}, {\overline{ijkl}}} \spc 
\, \Psi_{ijkl}(z)\spc \bar{\Psi}_{\overline{ijkl}}(\bz)
\\[2mm]
 \langle B^\dagger A^\dagger   B A \rangle \! \is \!  \sum_{{ijkl}, {\overline{ijkl}}} \spc  
 \Psi_{ijkl}(z^*)\spc \bar{\Psi}_{\overline{ijkl}}(\bz^*).
\label{otoc}
\eea
Here $\Psi_{ijkl}(z)$  and $\Psi_{ijkl}(z^*)$  are the respective chiral conformal blocks. 
The sum over all four pairs of indices runs over the full spectrum of the CCFT.  With a slight abuse of notation, we are using the the complex coordinate $z$ as short-hand for the location of the four operators along the celestial torus rather than just the cross ratio. The conformal blocks exhibits branch cuts for the special values of $z$ at which the two operators $A$ and $B$ are light-like separated.  The OTOC is obtained  from the TOC  by analytically continuing $z$ to a new value, which for brevity we denote by $z^*$. 
The conformal blocks on the celestial torus can be diagrammatically represented as 
\bea
\label{psitoc}
\Psi_{ijkl}(z)\! \is \! \delta_{ij} \, 
  \raisebox{-.5cm}{
\begin{tikzpicture}[scale=.63]
\draw[thick] (-1.6,0)--(-1,0)-- node[below]{$i$\ \,} (0,0) -- node[below]{$k$\, } (1,0) -- node[below]{\ $i$} (2,0) -- node[below]{\  $\ell$} (2.8,0);
\draw[thick] (-1.6,-1.3) --(2.8,-1.3);
\draw[thick]  (1.05,0) -- (1.05,.8) node[above]{$A$};
\draw[thick]  (2.03,0) -- (2.03,.8) node[above]{$B$};
\draw[thick]  (-1.1,0) -- (-1.1,.8) node[above]{$B^\dagger$};
\draw[thick]  (-.1,0) -- (-.1,.8) node[above]{$A^\dagger$};
\draw[thick]    (-1.6,-1.3) to[out=180,in=-90]  (-2.25,-.65) to[out=90,in=180] (-1.6,0);
\draw[thick]    (2.8,-1.3) to[out=0,in=-90] (3.5,-.65) to[out=90,in=0] (2.8,0);
\end{tikzpicture}
} \quad ; \quad
\Psi_{ijkl}(z^*) 
\, =\, 
  \raisebox{-.5cm}{
\begin{tikzpicture}[scale=.63]
\draw[thick] (-1.6,0)--(-1,0)-- node[below]{$i$\ \, } (0,0) -- node[below]{$k$\, } (1,0) -- node[below]{\ $j$} (2,0) -- node[below]{\ $\ell$} (2.8,0);
\draw[thick] (-1.6,-1.3) --(2.8,-1.3);
\draw[thick]  (1.05,0) -- (1.05,.8) node[above]{$B$};
\draw[thick]  (2.03,0) -- (2.03,.8) node[above]{$A$};
\draw[thick]  (-1.1,0) -- (-1.1,.8) node[above]{$B^\dagger$};
\draw[thick]  (-.1,0) -- (-.1,.8) node[above]{$A^\dagger$};
\draw[thick]    (-1.6,-1.3) to[out=180,in=-90]  (-2.25,-.65) to[out=90,in=180] (-1.6,0);
\draw[thick]    (2.8,-1.3) to[out=0,in=-90] (3.5,-.65) to[out=90,in=0] (2.8,0);
\end{tikzpicture}
}.  \\[-1mm] \notag
\eea
Here $i, j, k$ and $\ell$ label the intermediate Virasoro representations. 

In equation \eqref{psitoc}, we used the fact that the two pairs of operators $A^\dag(t_1)$, $A(t_1)$ and $B^\dag(t_0)$, $B(t_0)$ are each right on top of each other. In this limit, we can use the fact that  the identity operator gives the dominant contribution in the OPE between the two pairs of operators to make the replacement
\bea \label{crossing2}
\raisebox{-.65cm}{
\begin{tikzpicture}[scale=.65]
\draw[thick] (0,0)node[left]{$i$\ } -- (1,0) -- node[below]{$k$} (2,0) -- (3,0)  node[right]{$j$} ;
\draw[thick]  (.9,0) -- (.9,.95) node[above]{$A^\dag$};
\draw[thick]  (2.1,0) -- (2.1,.95) node[above]{$A$};
\end{tikzpicture}} \is
F^A_{ik} \; \delta_{ij}\spc  \raisebox{-.35cm}{
\begin{tikzpicture}[scale=.62]
\raisebox{-.1cm}{\draw[thick] (0.4,0)node[left]{$i$\ } -- (1,0) -- (2,0) -- (2.6,0)  node[right]{$i$} ;
\draw[dashed]  (1.5,0)  -- (1.5,.8) ;
\draw[thick]  (1.5,.8) -- (.9,1.4) node[left]{$A^\dag$\!};
\draw[thick]  (1.5,.8) -- (2.1,1.4) node[right]{\!$A$};} 
\end{tikzpicture}
} 
\eea
where the dotted line indicates the vacuum channel and $ F^A_{ik} \spc =F_k^0\bigl[\mbox{$\raisebox{-1pt}{\scriptsize${A\spc A}$}\atop{i\; i }$}\bigr]$ is an appropriate fusion matrix of the CFT. The fusion coefficients $F^A_{ik}$ are universal for Virasoro CFTs with given central charge and were computed by Ponsot and Teschner~\cite{Ponsot:1999uf}. We will quote a special limit of their result later on.  A similar equation holds for $B$ and $B^\dag$.
Combining the two relations, we find that the TOC conformal block simplifies to the relation
\bea
\label{psitoc2}
\raisebox{-.5cm}{
\begin{tikzpicture}[scale=.64]
\draw[thick] (-1.75,0)--(-1,0)-- node[below]{$i$\ \,} (0,0) -- node[below]{$k$\, } (1,0) -- node[below]{\ $i$} (2,0) -- node[below]{\ \ \ $\ell$} (2.8,0);
\draw[thick] (-1.75,-1.3) --(2.8,-1.3);
\draw[thick]  (1.05,0) -- (1.05,.8) node[above]{$A$};
\draw[thick]  (2.03,0) -- (2.03,.8) node[above]{$B$};
\draw[thick]  (-1.1,0) -- (-1.1,.8) node[above]{$B^\dagger$};
\draw[thick]  (-.1,0) -- (-.1,.8) node[above]{$A^\dagger$};
\draw[thick]    (-1.75,-1.3) to[out=180,in=-90]  (-2.4,-.65) to[out=90,in=180] (-1.75,0);
\draw[thick]    (2.8,-1.3) to[out=0,in=-90] (3.45,-.65) to[out=90,in=0] (2.8,0);
\end{tikzpicture}}
 \, 
 \is 
  F^A_{ik} \, F^B_{i\ell}\,
  \raisebox{-.5cm}{
\begin{tikzpicture}[scale=.64]
\draw[thick] (-1.75,0)--(-1,0)--  (0,0) -- node[below]{$i$\ \, } (1,0) -- (2,0) -- (2.75,0);
\draw[thick] (-1.75,-1.3) --(2.75,-1.3);
\draw[dashed]  (-1,0)  -- (-1,.8) ;
\draw[thick]  (-1,.75) -- (-.5,1.2) node[right]{\!$B$};
\draw[thick]  (-1,.75) -- (-1.5,1.2) node[left]{$B^\dag$\!}; 
\draw[dashed]  (2,0)  -- (2,.8) ;
\draw[thick]  (2,.75) -- (1.5,1.2) node[left]{$A^\dag$\!};
\draw[thick]  (2,.75) -- (2.5,1.2) node[right]{\!$A$};
\draw[thick]    (-1.75,-1.3) to[out=180,in=-90]  (-2.4,-.65) to[out=90,in=180] (-1.75,0);
\draw[thick]    (2.75,-1.3) to[out=0,in=-90] (3.4,-.65) to[out=90,in=0] (2.75,0);
\end{tikzpicture}}\\[-1mm]\notag
  \eea
representing the fact that ${\langle B^\dagger A^\dagger A B\rangle}$ factorizes into the product ${\langle B^\dagger B\rangle\langle A^\dagger A \rangle}$ of two point functions.

\subsubsection{OTOC from the Monodromy Matrix}  
\vspace{-1mm}
To compute the ratio \eqref{otocratio}, we need the ability to exchange operators. This can be done either by explicit analytic continuation of the relevant conformal blocks, or by means of the crossing matrices that implement the basis change between the two orderings. We first describe the latter method. The relevant monodromy properties of 
CCFT correlation functions can be studied by standard techniques of 2D conformal field theory.\footnote{While low point CCFT correlators defined through the standard celestial holographic dictionary have various exotic features, we will retain the optimistic assumption that these exotic features are artefacts of decomposing a Poincar\'e invariant theory into its Lorentz subgroup, and that these features will not obstruct the analytic continuation and monodromy properties of higher-point conformal blocks used in this section. } 

We can exchange the operator ordering using the crossing matrix relating conformal blocks associated to different channels of the four point correlator. The space of four-point conformal blocks is a linear space with different possible basis choices. The crossing matrices are the unitary basis transformations that relate two different bases corresponding to the different ways of summing over a complete sets of intermediate states. For our purpose, the relevant crossing operator is the one that interchanges the order of two operators
\bea \label{crossing}
\raisebox{-.55cm}{
\begin{tikzpicture}[scale=.7]
\draw[thick] (0.1,0)node[left]{$k$} -- (1,0) -- node[below]{$j$} (2,0) -- (2.9,0)  node[right]{$\ell$} ;
\draw[thick]  (.9,0) -- (.9,.85) node[above]{$B$};
\draw[thick]  (2.1,0) -- (2.1,.85) node[above]{$A$};
\end{tikzpicture}
} 
\is \, \sum_{i} \, R_{j}^{\spc i}\bigl[\mbox{${k \, A}\atop{\ell\, B }$}\bigr] \,
\raisebox{-.65cm}{
\begin{tikzpicture}[scale=.7]
\draw[thick] (0.1,0)node[left]{$k$} -- (1,0) -- node[below]{$i$} (2,0) -- (2.9,0)  node[right]{$\ell$} ;
\draw[thick]  (.9,0) -- (.9,.85) node[above]{$A$};
\draw[thick]  (2.1,0) -- (2.1,.85) node[above]{$B$};
\end{tikzpicture}
}.
\eea
Here $i, j, k$ and $\ell$ denote the Virasoro representations. The matrix $R_{j}^{\spc i}\bigl[\mbox{${k \, A}\atop{\ell\, B }$}\bigr]$  is called the R-matrix. In holographic terms, it represents the partial wave decomposition of the  2-particle scattering matrix between the bulk excitations created by the local CCFT operators $A$ and $B$.

Crossing matrices in CFT, like the R-matrix, are determined by the conformal representation theory of the Virasoro algebra with a given central charge. An explicit expression of the R-matrix of Virasoro CFT is given in~\cite{Ponsot:1999uf}. We will not write the explicit result here, except to note that for our purpose we should take the limit of large imaginary central charge $c$. In this limit, the crossing matrix can be expressed in terms of the 6j-symbol of the 2D global conformal group, which (due to the imaginary value of $c$) we should identify with SU$(1,1)$. This specific large $c$ limit happens to be the same one that reduces the 2D CFT correlations functions and monodromy matrices to those of 1D Schwarzian quantum mechanics. The following discussion directly borrows from \cite{Mertens:2017mtv} and \cite{Lam:2018pvp}. 

The analytic continuation that relates the time-ordered conformal block $\Psi_{ijkl}(z)$ and the out-of-time-ordered conformal block $\Psi_{ijkl}(z^*)$ involves moving one operator past the light cone of the other. So the point $z^*$ lies on the second sheet of the associated branch cut. 
 To obtain the linear relationship between the basis of conformal blocks evaluated on the first and second sheet, it is sufficient to apply the local R-matrix relation \eqref{crossing} associated to the interchange of the two operators $A$ and $B$. In diagrammatic notation, the crossing relation reads
\bea
\label{tcrossing}
  \raisebox{-.66cm}{
\begin{tikzpicture}[scale=.66]
\draw[thick] (-1.75,0)--(-1,0)-- node[below]{$m$\ \; } (0,0) -- node[below]{\ $k$} (1,0) -- node[below]{\ $j$} (2,0) -- node[below]{$\ell$\ } (3,0);
\draw[thick] (-1.75,-1.3) --(3,-1.3);
\draw[thick]  (1.05,0) -- (1.05,1) node[above]{$B$};
\draw[thick]  (2.03,0) -- (2.03,1) node[above]{$A$};
\draw[thick]  (-1.15,0) -- (-1.15,1) node[above]{$B^\dagger$};
\draw[thick]  (-.2,0) -- (-.2,1) node[above]{$A^\dagger$};
\draw[thick]    (-1.75,-1.3) to[out=180,in=-90]  (-2.4,-.65) to[out=90,in=180] (-1.75,0);
\draw[thick]    (3,-1.3) to[out=0,in=-90] (3.65,-.65) to[out=90,in=0] (3,0);
\draw[magenta] (.25,-.7)--(3,-.7) -- (3,1.7) --(.25,1.7) -- (.25,-.7);
\end{tikzpicture}
}
 \is 
 \; \sum_{i} R^{\spc i}_{{j}}\bigl[\mbox{${k \, A}\atop{\ell\, B }$}\bigr]~
   \raisebox{-.66cm}{
\begin{tikzpicture}[scale=.66]
\draw[thick] (-1.75,0)--(-1,0)-- node[below]{$m$\ \;} (0,0) -- node[below]{\ $k$} (1,0) -- node[below]{\ $i$} (2,0) -- node[below]{$\ell$\ } (3,0);
\draw[thick] (-1.75,-1.3) --(3,-1.3);
\draw[thick]  (1.05,0) -- (1.05,1) node[above]{$A$};
\draw[thick]  (2.03,0) -- (2.03,1) node[above]{$B$};
\draw[thick]  (-1.15,0) -- (-1.15,1) node[above]{$B^\dagger$};
\draw[thick]  (-0.2,0) -- (-0.2,1) node[above]{$A^\dagger$};
\draw[thick]    (-1.75,-1.3) to[out=180,in=-90]  (-2.4,-.65) to[out=90,in=180] (-1.75,0);
\draw[thick]    (3,-1.3) to[out=0,in=-90] (3.65,-.65) to[out=90,in=0] (3,0);
\draw[magenta] (.25,-.7)--(3,-.7) -- (3,1.7) --(.25,1.7) -- (.25,-.7);
\end{tikzpicture}
}.\\[-1mm]\notag
\eea
The R-matrix only acts on the Virasoro representation label of the intermediate channel between the two operators $A$ and $B$ that need to be exchanged in going from the TOC to the OTOC. We furthermore have made use of the fact that $\Psi_{ijkl}(z)$ contains a factor of $\delta_{ij}$ to collapse the sum in \eqref{crossing} to a single term.

We can now compute the OTOC  ratio \eqref{otocratio} as follows. First we decompose the OTOC into conformal blocks. Then we move the operators $A$ and $B$ to the same time instant. This produces a simple time evolution phase $e^{ih_j t}$, 
where $t = t_1-t_0$ is the time-separation between the $A(t_1)$ and $B(t_0)$ operator insertions and $h_j$ the conformal dimension of the $j$ channel. 
 Next we apply the crossing relation \eqref{tcrossing}. Finally, we move the operators $A$ and $B$ back to their original time instants by including a phase $e^{-i h_i t}$. This yields the following result of the OTOC  ratio 
  \bea
\label{otocaa}
\frac{\langle B^\dagger A^\dagger B A \rangle }{\langle B^\dagger B\rangle\langle A^\dagger A \rangle}\is  \sum_{{ijkl}, {\overline{ijkl}}} \spc 
{\cal A}_{ijkl}(t)\;\;  
\overline{\!{\cal A}_{ijkl}( t) }
\eea
where\\[-8mm] 
\bea
\label{otocamp}
{\cal A}_{ijkl}(t) 
  \is  e^{i(h_j-h_i) t }   R_{{i}{j}}\bigl[\mbox{${k \, A}\atop{\ell\, B }$}\bigr] \,  F^A_{ik}\spc F^B_{i\ell}\\[-4mm] \notag
\eea
and similar expression holds for $\,\overline{\!{\cal A}_{ijkl}(t)\!}\,$. Here $(h_i,\bar{h}_i)$ and $(h_j,\bar{h}_j)$ are the left and right conformal dimensions of the intermediate $i$ and $j$ channel. 

The discussion so far has been very general and admittedly somewhat abstract. The pay-off, however, is that by plugging in the known results for the crossing matrices of the Virasoro CFT, equations \eqref{otocaa}-\eqref{otocamp} immediately give us practical explicit expression for the chiral components of the OTOC. The result further simplifies by virtue of the fact that the CCFT has a divergent central charge. As mentioned above, the CFT crossing matrices in this limit reduce to those of Schwarzian quantum mechanics, and can be expressed in terms of the Clebsch-Gordan and 6j-symbols of the global conformal group SU$(1,1)$. The relevant calculations are described in detail in \cite{Mertens:2017mtv} and \cite{Lam:2018pvp}. Here we will just quote the result. 

It will be convenient to introduce the notation
\bea
h_k\nspc - h_j\spc  =\spc {i\nu_1}, \qquad  h_j \nspc - h_\ell \spc = \spc {i\nu_2}, \qquad h_k \nspc - h_i \spc = \spc {i\nu_3} \qquad h_i  - h_\ell \spc = \spc {i\nu_4}.
\eea
One can think of each $\nu_i$ as the left-moving energy injected by each of the four operators into the correlator. We further make the simplifying assumption that the operator $A$ and $B$ both have the same conformal dimension $h$. The chiral OTOC amplitude then reads as follows
\bea
\label{asmat}
{\cal A}_{ijkl}(t) \is e^{{i}(\nu_3\nspc  - \nu_1\nspc )t}\; 
\, {\langle \nu_4,\nu_3 | \, {\cal S}\, |\nu_2,\nu_1\rangle } \\[-7mm]\notag
  \eea
where \\[-7mm]
\bea
\label{smat1}
\langle \nu_4,\nu_3 | {\cal S} |\nu_2,\nu_1\rangle\!\! \is \! {(4\pi i \epsilon)^{i (\nu_{1}\nspc -\nu_3)}}\,  \Bigl[\prod_{a=1}^4 e^{\pm \frac \pi 2 \nu_a } \textstyle \Gamma(h\pm i \nu_a) \Bigr]\, \textstyle \Gamma( i (\nu_{1}\!-\nspc\nu_3))
\eea
times the usual energy conservation delta function $2\pi  \delta(\nu_{1}\!+\!\nu_{2}  \! -\! \spc \nu_{3}\! -\nu_4)$. 

The notation of the OTOC chiral amplitude as an $\mathcal{S}$-matrix element is deliberate. We can rewrite the right-hand side of \eqref{smat1} as an overlap integral 
of the following gravitational shockwave $\mathcal{S}$-matrix
\bea
\label{thooft}
{\cal S} \is e^{4\pi i \epsilon p_+p_-}
\eea
between four 2D Rindler mode functions. Here $p_+$ and $p_-$ represent the Minkowski light-cone momenta and the $\nu_a$ are Rindler energies. 
Equation \eqref{thooft} is the 2D 't Hooft $\mathcal{S}$-matrix~\cite{Dray:1984ha} that encodes the gravitational shift $x^- \to x^- + 4\pi \epsilon p_+$ on a right moving trajectory due the presence of a left-moving particle with lightcone momentum $p_+$. This shift has an exponentially growing effect when viewed in Rindler coordinates. This is a first hint of Lyapunov behavior in CCFT. 

In the next section we will re-derive the above result via the well-tested assumption that 2D CFT correlation functions in the large $c$ limit are dominated by a suitably chosen vacuum conformal block.

\subsubsection{OTOC from the Vacuum Block}  
\vspace{-1mm}

The above treatment of the OTOC conformal blocks only made use of the Virasoro symmetry of CCFT. The emergence of gravitational dynamics from this subsector is not surprising, given its close relationship with AdS${}_3$ gravity. If we want to say more about the OTOCs, we would need to know about and use more of the specific properties of the spectrum, fusion rules, OPE coefficients and extended symmetries of CCFT. These more detailed properties are all implicitly contained in the sum over the intermediate channels in \eqref{otocaa}. 

Gravitational saddle points describe universal or appropriately averaged properties of holographic CFTs. One practical implementation of this philosophy is that gravitational saddle points can often be identified with the contribution of an appropriate vacuum conformal block. The time-ordered correlation function can indeed be argued to be given by an identity conformal block
\bea
\label{tocpsipsi}
\frac{\langle B^\dag_4 A^\dag_3 A_2 B_1\rangle }{\langle A^\dag_3 A_2\rangle\langle B_4^\dag B_1\rangle}\is  {\Psi} \Bigl( {}^{A}_{A} \; {}^{B}_{B} , {\rm vac},z\Bigr)\;\, \overline{\!{\Psi} \Bigl( {}^{A}_{A} \; {}^{B}_{B} , {\rm vac},z\Bigr)\!}
,~~~~~~~~~~z= -\frac{\sinh \frac 1 2 t_{23} \spc \sinh\nspc \frac 12 t_{14} }{\sinh\nspc \frac 12  t_{12} \spc \sinh\nspc \frac 12 t_{34}}.
\eea
Here the vacuum block is defined on the sphere, or equivalently, the projective plane, and $z$ denotes the cross ratio of the coordinate location of the four operators on the sphere.  The planar vacuum block dominates for two reasons. First, as before, we assume that the two pairs of operators $A^\dag$, $A$ and $B^\dag$, $B$ are pairwise very close to each other. We already used this above to write equation \eqref{psitoc2}. Second, we can use an exponential conformal mapping to unwind the thermal circle and re-express a thermal correlation function as a vacuum expectation value at zero temperature, but with an exponential identification of coordinates. 

The reasoning that the vacuum block dominates for certain correlation functions generalizes Cardy's argument for determining the high temperature behavior and asymptotic spectrum of CFTs through vacuum block dominance of the torus partition function. Comparing \eqref{tocpsipsi} with equation \eqref{toc}, we see that, just as in the case of the Cardy spectrum, the sum over intermediate sectors simply factorizes into two independent sums. Moreover, we learn that the spectral properties of the CCFT should be such that the sum produces a chiral vacuum block. The above physical argumentation is not rigorous, but is well motivated for the case of standard holographic 2D CFTs in the context of the AdS/CFT correspondence. Let us assume the the same reasoning can be applied to CCFT. 

The out of time ordered correlation function is obtained by analytically continuing the time ordered conformal blocks to the second sheet. Assuming the vacuum blocks in \eqref{tocpsipsi} continue to provide the dominant contribution after the analytic continuation, we deduce that
\bea
\frac{\langle B_4^\dag A_3^\dag B_1 A_2\rangle }{\langle A_3^\dag A_2\rangle\langle B_4^\dag B_1\rangle}\is {\Psi} \Big( {}^{A}_{A} \; {}^{B}_{B} , {\rm vac},z^*\Big)\;\, \overline{\!{\Psi} \Big( {}^{A}_{A} \; {}^{B}_{B} , {\rm vac},z^*\Big)\!}.
\eea
The vacuum conformal block at large central charge is explicitly known \cite{Chen:2016cms}
\bea
\label{block}
\lim_{\raisebox{-3.5pt}{\footnotesize ${{c\to \infty}\atop {x=cz~{\rm fixed}}}$}} {\Psi} \Big( {}^{A}_{A} \; {}^{B}_{B} , {\rm vac},z^*\Big) \!\is \! x^{-2h} U(2h,1,1/x),
\qquad \ \ 
{x}\, =  \, \frac{i}{4\pi \epsilon}\, \frac{e^{\frac 1 2 (t_1+t_2 - t_3-t_4)}}{4\sinh \nspc \frac 1 2 t_{12} \spc \sinh\nspc  \frac 1 2 t_{34}}
\eea
in terms of the confluent hypergeometric function, defined as the integral $U(a,1,y) = \frac{1}{\Gamma(a)} \int_0^\infty\! ds\, e^{-sy} \frac{s^{a-1}}{(1+s)^a}$.
Combining the two chiral blocks gives the following explicit result for the OTOC
\bea
\frac{\langle B^\dag_4 A^\dag_3 B_1 A_2\rangle }{\langle A^\dag_3 A_2\rangle\langle B_4^\dag B_1\rangle}\! \is \! x^{-2h}\bar{x}^{-2\bar{h}} \, U(2h,1,1/x) \;  U(2\bar{h},1,1/\bar{x}).
\eea

Some brief remarks are in order. First, the explicit expression \eqref{block} of the chiral OTOC follows  by integrating the result \eqref{asmat}-\eqref{smat1} for the chiral OTOC conformal block with uniform measure over all frequencies $\nu_a$
\bea  
\Bigl[\prod_{a=1}^4 \int\! 
\frac{d\nu_a}{2\pi} e^{\pm i\nu_a t_a}  \Bigr]\, 
\, {\langle \nu_4,\nu_3 | \, {\cal S}\, |\nu_2,\nu_1\rangle }  \is \! x^{-2h} \, U(2h,1,1/x).
\eea
This suggests that the spectrum of the CCFT in the regime of interest is well approximated by the usual Cardy spectrum. 
Secondly, this result matches the OTOC in Schwarzian QM. In nAdS${}_2$ holography, it describes the scattering of two particles that collide in the proximity of the horizon of the 2D black hole in JT gravity. It in particular exhibits the anticipated maximal Lyapunov behavior.  

\subsubsection{OTOC from Celestial Backreaction} 
\vspace{-1mm}

Finally, we present a geometric derivation of the OTOC based on 4D physics and our earlier description in section 3 of the backreaction due to the insertion of local CCFT operators on the celestial sphere. This gravitational derivation of the OTOC looks a priori quite different from the above more technically sophisticated CFT analysis, but the two are both directly linked via the so-called monodromy method for determining the explicit form and monodromy properties of 2D conformal blocks.

As explained in section 4, the exponential behavior of CCFT correlation functions has a simple geometric origin in terms of the coordinate identification from the celestial sphere to the celestial torus 
\bea (z,\bar{z}) \!\is\! (e^{i(\tau_E+\phi)},e^{i(\tau_E-\phi)}), \qquad {\rm with} \qquad
\tau_E \,=\, \psi + i t\,
\eea 
 the complexified Rindler time coordinate. After Wick rotating, this coordinate relation shows that evolution in the Lorentzian Rindler time coordinate $t$ describes an exponential approach towards the origin of the $(z,\bar{z})$ plane
\bea 
\label{tshrink}
(z,\bar{z}) \! \is \! (e^{-t+i\phi},e^{-t-i\phi}). 
\eea
We immediately see that any small backreaction in the form of an infinitesimal coordinate shift in $z$ would cause an exponentially growing Shapiro time delay as measured in the Rindler time coordinate $t$. This is an expected consequence of the fact that the late time Rindler observer is exponentially close to the Rindler horizon, and  thus correspondingly sensitive to infinitesimal shifts relative to the location of the horizon.

In the following we will exhibit the butterfly effect caused by a local CCFT operator $B(z_1)$.   As explained in section~\ref{sec:vir}, the stress energy associated with this local operator induces a small geometrical defect in the form of an infinitesimal angle deficit around the location $z=z_1$. This angle deficit can be incorporated by means of the infinitesimal coordinate transformation (here, for simplicity, we only write the holomorphic part of the transformation)
\bea
\label{cdefect}
B(z_1): \qquad 1 - \frac z{\raisebox{1pt}{$z_1$}} \! & \to & \!  \Bigl(1- \frac z{\raisebox{1pt}{$z_1$}}\Bigr)^{1-2i\epsilon h_b}, \qquad \qquad -i\epsilon = \frac{3}{c}
\eea
where $h_b$ denotes the left-moving scale dimension of $B$ and $\epsilon$ the IR cut-off parameter. 

The above infinitesimal defect is the celestial imprint of the full 4D backreaction associated with the local CCFT operator.
The resulting butterfly effect on another local operator $A$ is encoded in the out-of-time correlation function, or equivalently, in the expectation value of the commutator squared $
\bigl\langle[A(z_2), B(z_1)]^2\bigr\rangle$~\cite{Roberts:2014ifa,Jackson:2014nla}.
At late times, the leading contribution to this commutator squared comes from considering the effect of the coordinate shift \eqref{cdefect} induced by the insertion of $B(z_1)$ on the operator $A(z_2)$ as it approaches the origin. The effect of this shift becomes visible by performing a monodromy transformation by moving the operator $A(z)$ around the location of $B(z_1)$, or equivalently, by analytically continuing the correlation function to the second sheet. A simple calculation gives that in the small $z_2$ limit
\bea
A(z_2) B(z_1) \! \is \! B(z_1)A(z_2)\bigr|_{{\rm 2nd\, sheet }}
\! \simeq \,  B(z_1) A(z_2\! +\nspc 4\pi i \epsilon h_b z_1).
\eea
Inserting \eqref{tshrink}, combining the left- and right parts and considering only $t$ dependence, we find that
\bea\label{commutator}
[A(t_2), B(t_1)] \! & \! \simeq \! &  \! 2\pi i \epsilon \Delta_b\cdot e^{t_2-t_1}B(t_1)\, \partial_{t_2} A(t_2)\, 
\eea
with $\Delta_b$ the full scale dimension of $B(z_1)$.

Equation \eqref{commutator} expresses the gravitational backreaction due to the $B$ operator on the location of the $A$ operator. However, the situation is symmetric: the operator $A$ also creates a geometric defect that shifts the location of the operator $B$. We can write the commutation relation in a more suggestive and symmetric form by noting that $\Delta_b$ is the energy of the state created by the operator $B$. Hence for early $t_1$ we can use the state operator correspondence to equate $\Delta_b B(t_1) =  \partial_{t_1} B(t_1)$. The above commutation equation then becomes 
\bea\label{commutatortwo}
[A(t_2), B(t_1)] \! & \! \simeq \! &  \! 2\pi i \epsilon\,  e^{t_2-t_1}\partial_{t_1} B(t_1)\, \partial_{t_2} A(t_2).\, 
\eea
Again we see that the exponential growth of the OTOC is caused by a geometric shockwave interaction.

\section{Conclusion}
\vspace{-1mm}
In this paper we have argued that celestial conformal field theory, when viewed as a dynamical quantum system with unitary Hamiltonian time evolution, exhibits characteristics  of maximal quantum chaos. To build our case, we re-examined the soft phase space associated to the superrotation symmetry of the 4D space-time and used the presence of the (2,0) Goldstone current~\cite{Ball:2019atb} to  introduce two celestial stress tensors that generate two mutually commuting Virasoro algebras with a divergent imaginary central charge.
Restricting to operators that commute with one of the Virasoro algebras leads to backreaction effects which can be most clearly brought to light by means of the out-of-time-ordered correlators. We studied the OTOCs and demonstrated the Lyapunov growth using standard 2D CFT technology for large-$c$ systems. 

The physical origin of this chaotic behavior lies in the identification of the time coordinate of CCFT, defined through radial quantization, with the Rindler time coordinate experienced by an accelerating observer in 4D space-time. Relating the celestial correlators to the observation of this accelerating observer involves an analytic continuation from the celestial sphere to the celestial torus.   The celestial torus perspective naturally incorporates the fact that the CCFT dynamics takes place at finite temperature, in turn, matching the bulk interpretation in terms of the Rindler observer.  This perspective also illuminates the appearance of two independent Virasoro algebras as the asymptotic symmetry groups of the two AdS caps that meet at the celestial torus, which ties back into our Goldstone mode analysis. 

We are lead to the following natural future directions and open questions:

\vspace{.5em}

\noindent {\it Gluing Construction --} Interpreting our doubled-Virasoro algebra in terms of two large-radius AdS caps meeting at the celestial torus suggests a natural generalization to other vacuum transitions and currents.  On the bulk gravitational side we have the impulsive wave analyses of~\cite{Nutku,Compere:2019odm,Freidel:2021qpz}.  From the celestial CFT current algebra we have the $w_{1+\infty}$ symmetry of~\cite{Guevara:2021abz,Strominger:2021lvk}.  The (A)dS$_3$ picture advocated here and in~\cite{deBoer:2003vf,Cheung:2016iub,Ball:2019atb} presents a natural route to toy examples of a CCFT and bulk dual pair that captures this symmetry algebra. We will further explore this tantalizing prospect in~\cite{PV3}.  

\vspace{.5em}

\noindent {\it Incorporating Translations -- } In this paper we have focused on the dynamical properties of CCFT that follow from superrotation symmetry.  Since Lorentz transformations and superrotations both act relative to a specific space-time point, translation and supertranslation symmetries are non-linearly realized in the celestial basis via spectrum shifting operators.  Understanding the relationship between our study and the supertranslation current would be of interest for several reasons, in particular because supertranslations are naturally linked to the standard 4D gravitational shock wave $\mathcal{S}$-matrix. The role of the $\epsilon$ deformation of CCFT is also of interest and may illuminate the connection between Celestial CFT and flat space limits of AdS~\cite{Hijano:2019qmi,Hijano:2020szl,Compere:2019bua}.

\vspace{.5em}

\noindent {\it Adding Horizons -- } Understanding how black hole physics is encoded in celestial CFT is an interesting open problem. As seen in~\cite{Pasterski:2020xvn}, extra boundary components introduce an enhancement in the soft phase space.   The observations in~\cite{Strominger:2016wns,Pasterski:2020pdk,Crawley:2021auj} point towards some natural starting points in terms of limits of scattering amplitudes or changing the modular parameter for CCFT on the celestial torus. 

\vspace{.5em}

We see that in the course of examining the conformally soft sector, we are confronted with and are able to gain insight into foundational open questions about CCFT.  The fact that we can predict and analyze the chaotic gravitational bulk dynamics, even without detailed knowledge of the full structure of CCFT, illustrates the power of 2D conformal symmetry when it comes to exhibiting interesting bulk physics and dynamics.

\subsection*{Acknowledgements}
\vspace{-1mm}

We thank Scott Collier, Laurent Freidel, Matthew Heydeman, Andrea Puhm,  Andrew Strominger, Joaquin Turiaci, Emilio Trevisani, Erik Verlinde, and Sasha Zhiboedov for useful discussions and comments. The research of SP is supported by the Sam B. Treiman Fellowship at the Princeton Center for Theoretical Science.  The research of HV is supported by NSF grant number PHY-1914860.

\appendix

\section{Local operators on 
\texorpdfstring{$\mathbb{T}^{1,1}$}{the celestial torus}}\label{sec:TOp}
\vspace{-1mm}

In this appendix we describe the map between 4D operators ${\mathcal O}(X)$ and CCFT operators on the celestial torus.  We review the salient features of~\cite{Atanasov:2021oyu}, and emphasize the connection to  the construction described in section~\ref{sec:dictionary}. A practical way to deal with this is to first construct the CCFT operators that live on the intersection between the celestial sphere and the celestial torus. At this special locus, the CCFT operators can be imported from the (1,3) signature space-time by means of the Klein-Gordon overlap with the corresponding space-time wavefunction. 

We can describe the intersection locus between $\mathbb{S}^2$ and $\mathbb{T}^{1,1}$ as follows. Plugging the parametrization \eqref{complexz} of the celestial coordinates into formula \eqref{ptosphere} for the reference direction $q^\mu$ gives
\bea
\label{qgen}
q^\mu(t,\psi,\phi) 
\! \is\! e^{i\psi} {p}^\mu(t,\psi,\phi)\, =\, e^{-t+i\psi}\bigl(\cosh(t\nspc- \nspc i\psi),\cos\phi,\sin\phi,\sinh(t\nspc -\nspc i\psi)\bigr).
\eea
After Wick rotating $X_3$, the celestial torus is the real time slice $
\mathbb{T}^{1,1} =
\{ t=0\}$. The $(2,2)$ four vector $\hat{q}^\mu = e^{i\psi} \hat{p}^\mu$ then reduces to 
\bea 
\label{qtorus}
\mathbb{T}^{1,1}: \qquad \hat{p}^\mu(0,\psi,\phi) 
\! \is \! 
\bigl(\cos\psi,\cos\phi,\sin\phi, \sin \psi\bigr).
\eea
The celestial sphere is obtained either by setting $\psi=0$ or by setting $\psi=\pi$ in \eqref{qgen}. Hence the celestial torus and the two celestial spheres intersect along the two equatorial circles $S^1_+\! = \bigl\{t\nspc =\nspc 0,\psi\nspc =\nspc 0\bigr\}$ and $
S^1_-\! = \bigl\{ t\nspc  =\nspc  0,\psi\nspc = \nspc \pi \bigr\}$
along which the momenta are both real
\bea
S^1_+ : \qquad 
\hat{p}^\mu(0,0,\phi)\!\! \is\!   (\spc 1, \smpc\cos\phi\smpc,\smpc \sin\phi\smpc,\smpc 0 \spc ),\\[2mm] S^1_-: \qquad 
\hat{p}^\mu(0,\pi, \phi 
)\!\! \is\!\!   (-1, \cos\phi, \sin\phi, 0).
\eea
Note that $\hat{p}^\mu = - \hat{q}^\mu$ along $S^1_-$.  We will also consider the two special points $P_+\! = \bigl\{ t=\nspc \infty,  \psi = 0\bigr\}$ and $P_-\! = \bigl\{ t = \infty,\spc \psi = \pi\bigr\}$ corresponding to the special light-like four momenta 
\bea
P_+ : \qquad {q}^\mu(-\infty,0,\phi)\!\! \is \textstyle \frac 1 2 \bigl(\spc 1\smpc ,\smpc 0\smpc ,\smpc 0\smpc,\smpc 1\spc \bigr),\\[2mm] P_-: \qquad  {q}^\mu(-\infty,\pi, \phi
)\!\! \is\!\!   -\textstyle \frac 1 2 \bigl(1, 0, 0, 1\bigr).
\eea
Both points map to  the same reference direction for the momentum vector $q^\mu = e^{i\psi} p^\mu =  \frac 1 2 (1,0,0,1)$ corresponding to the coordinate origin $(z,\bz) = (0,0)$ at the north pole of the celestial sphere. We~will interpret $P_\pm$ as the center points of the two equatorial circles $S^1_\pm$. The two circles and their center points will play a role in the construction of the local operators and  the Hilbert space of the CCFT  on ${\mathbb T}^{1,1}$.

Let us now proceed with the construction of local operators on~$\mathbb{T}^{1,1}$ starting from primary wavefunctions in (2,2) signature as in~\cite{Atanasov:2021oyu}. We will stick to the scalar case for simplicity. The space-time wavefunctions corresponding to the scalar primaries on $\mathbb{T}^{1,1}$ take the form
\bea\label{pvq}
\hat\Phi_{\Delta} (X;\psi,\phi)\!\! \is \!\! \frac{1}{(-\hat{p}\cdot\nspc X)^\Delta\!} \qquad {\rm with} \qquad \hat{p} = (\cos\psi, \cos \phi, \sin \phi,  \sin \psi)
\eea
where the reference four vector \eqref{qtorus} is labeled by the point ($\psi,\phi$) on the celestial torus.  
In constructing operators from these wavefunctions, we run into the technical subtlety that there is no obvious analog of the Klein-Gordon inner product \eqref{kgin} in (2,2) signature, that can be used to project the operator ${\mathcal O}(X)$ along  the space-time wavefunction~\eqref{pvq}. Namely, in $\mathbb{R}^{1,3}$ the in and out wavefunctions are prepared by an $i\epsilon$ prescription that regulates the Mellin transform
 \bea \hat\Phi^\pm_{\Delta} (X)\! \is \! \hat\Phi_{\Delta} (X_\pm)
 \eea where $X_\pm^0=X^0\mp i\epsilon$. While there is no notion of time ordering in $\mathbb{K}^{2,2}$, we can still analytically continue $X_\pm$ as in the $\mathbb{R}^{1,3}$ case on the $\psi=\{0,\pi\}$ loci. We find
\bea
\hat{\mathcal{O}}^\pm_{\Delta}(0,\phi)\!\! \is \!\! i\bigl(\smpc \hat{O}\smpc ,\smpc \hat{\Phi}^\pm_\Delta(0,\phi)\spc \bigr)_{\Sigma}, \qquad \qquad 
\hat{\mathcal{O}}^\pm_{\Delta}(\pi,\phi)\spc = \spc  i\bigl(\smpc \hat{O}\smpc ,\smpc \hat{\Phi}^\pm_\Delta(\pi,\phi)\spc \bigr)_{\Sigma}.
\eea
Comparing to the primaries constructed in section~\ref{sec:dictionary}, we see from~\eqref{qgen} that
\bea
\hat{\mathcal{O}}^\pm_{\Delta}(0,\phi)={\mathcal{O}}^\pm_{\Delta}(0,\phi),~& &~ \hat{\mathcal{O}}^\pm_{\Delta}(\pi,\phi)=- e^{-\pi\lambda}\hat{\mathcal{O}}^\pm_{\Delta}(0,\phi+\pi).
\eea
These operators sit at the two cuts illustrated in figure~\ref{torustfd}. We see an overcompleteness originating from how we've continued to use a $\pm i\epsilon$ prescription when we've analytically continued $X^\mu$ to $\mathbb{K}^{2,2}$.   The space of independent in and out single particle states on the celestial equator can be mapped to one or the other of the cycles.

We can then use the symmetry generators to translate these operators to generic points on $\mathbb{T}^{1,1}$. Continuing the $t$ parameter as in~\eqref{complexz} and taking into account the expected conformal factor between the sphere wavefunctions~\eqref{deltaj2} and torus wavefunctions~\eqref{pvq} we have
\bea
U(-i\psi)\,\hat{\mathcal{O}}^\pm_{\Delta}( 0,\phi)\, U^{\dag}(-i\psi)\, =\,\hat{\mathcal{O}}^\pm_{\Delta}( \psi,\phi).
\eea
Comparing this to the bulk interpretation as evolution in the complexified Rindler time indeed reaffirms the thermofield double interpretation of~\eqref{torustfd}. 
Introducing the light cone coordinates $y^\pm=\psi\pm \phi$, these local operators on $\mathbb{T}^{1,1}$ are termed $H$-primaries in~\cite{Atanasov:2021oyu}, because they behave as highest weight operators under the following redefinition of the global Lorentz generators
\be\badat{3}
H_0(y)&=\frac{1}{2}(e^{i{y}^+}L_1-e^{-iy^+}L_{-1}),~~~H_{\pm 1}(y)&=iL_0\mp\frac{i}{2}(e^{i{y}^+}L_1-e^{-i{y}^+}L_{-1})
\\
\bar{H}_0(y)&=\frac{1}{2}(e^{i{y}^-}\bar{L}_1-e^{-i{y}^-}\bar{L}_{-1}),~~~\bar{H}_{\pm 1}(y)&=i\bar{L}_0\mp\frac{i}{2}(e^{i{y}^-}\bar{L}_1-e^{-i{y}^-}\bar{L}_{-1})\\
\eadat\ee
which also obey the standard SL$(2,\mathbb{R})\times SL(2,\mathbb{R})$ algebra
 \be
 [H_n,H_m]=(n-m)H_{n+m},~~~[\bar{H}_n,\bar{H}_m]=(n-m)\bar{H}_{n+m}.
 \ee
 We further note that by appropriately smearing these local $H$-primaries we can produce $L$-primaries that, as for $\mathbb{S}^2$ primaries inserted at the north pole, diagonalize $\{L_0,\bar{L}_0\}$  and are annihilated by $\{L_1,\bar{L}_1\}$.  In terms of the wavefunctions, we have the relation~\cite{Atanasov:2021oyu}
\be
\frac{2^\Delta}{(X^0-X^3)^{\Delta}}=\frac{1}{(2\pi)^2}\int_0^{2\pi} dx^+\int_0^{2\pi} dx^- e^{ih(x^++x^-)}\frac{1}{(-\hat{p}\cdot{X})^\Delta}
\ee
where $h=\bar{h}=\frac{1}{2}\Delta.$ We can apply the same same smearing to our local operators
and reproduce the states.

\bibliographystyle{utphys}
\bibliography{cCFT}

\end{document}